\documentstyle[11pt,aps,prl,preprint]{revtex}
\begin{document}
\draft
\title{Spectral Properties of the Anderson Impurity Model: 
Comparison of Numerical Renormalization Group and 
Non--Crossing Approximation}
\author
{T. A. Costi$^{1}$, J. Kroha$^{2}$ and P. W\"{o}lfle$^{1}$}
\address
{$^{1}$ Universit\"{a}t Karlsruhe, Institut f\"{u}r Theorie 
der Kondensierten Materie, 76128 Karlsruhe, Germany}
\address
{$^{2}$ Cornell University, LASSP, Ithaca, N.Y. 14853, U.S.A.}
\maketitle
\begin{abstract}
A comparative study of the numerical renormalization group and 
non--crossing approximation results for the spectral functions 
of the $U=\infty$ Anderson impurity model is carried out. The 
non--crossing approximation is the simplest conserving approximation and
has led to useful insights into strongly correlated models of magnetic
impurities. At low energies and temperatures the method is known to be
inaccurate for dynamical properties due to the appearance of 
singularities in the physical Green's functions. The problems in 
developing alternative reliable theories for dynamical properties 
have made it difficult to quantify these inaccuracies.
In this paper we show, by direct comparison with essentially
exact numerical renormalization group calculations for the auxiliary and 
physical particle spectral functions, that the main source of error in
the non--crossing approximation is in the lack of vertex corrections in
the convolution formulae for physical Green's functions.
We show that the dynamics of the auxiliary particles within NCA is
essentially correct for a large parameter region, including the physically
interesting Kondo regime, for all energy scales down to $T_{0}$, the low
energy scale of the model and often well below this scale. Despite the
satisfactory description of the auxiliary particle dynamics,
the physical spectral functions are not obtained accurately on scales
$\sim T_{0}$. Our results
suggest that self--consistent conserving approximations which include
vertex terms may provide a highly accurate way of dealing with strongly
correlated systems at low temperatures.
\end{abstract}
\section{Introduction}
In recent years the physics of strongly correlated Fermi systems
has attracted wide interest, in particular in the context of high
$T_c$ superconductors \cite{anderson.87} and their normal state properties 
\cite{fascinating-ideas}. Despite intense efforts, systematic and  
controlled theoretical methods for dealing with such systems
are  still lacking. At the heart of the problem is the effect of a strong
on--site Coulomb repulsion $U$ on fermions living on a lattice, in
particular in $d=2$ dimensions. At low energy such models can be 
mapped onto effective Hamiltonians with the constraint of no double
occupancy of any lattice sites. The projection onto the corresponding
subspace of Hilbert space is presumably responsible for a number of
unusual properties, most remarkably the spin--charge separation and 
non--Fermi liquid behaviour. Unfortunately the local constraints on
the site occupancy are difficult to handle. One possible formulation 
\cite{barnes.76}
uses auxiliary particles such as slave-bosons ($b$) and pseudo-fermions
($f_{\sigma}$) to represent an electron as a composite particle consisting
of a pseudo-fermion particle and a slave-boson hole ($f_{\sigma}^{\dagger}b$).
The local constraint is holonomic in this formulation, and is given by
the condition that the sum of the occupation numbers of pseudo-fermions
and slave-bosons is equal to unity at any lattice site, 
$\sum_{\sigma}n_{f\sigma}+n_{b}=1$. The constraint is closely related
to the local gauge symmetry of the system with respect to simultaneous 
$U(1)$ gauge transformations of $b$ and $f_{\sigma}$. It may be shown
that starting with certain mean field theories the constraint and its
consequences lead to the appearance of longitudinal and transverse 
fictitious gauge fields coupling equally to pseudo-fermions and
slave-bosons\cite{baskaran.88}. The gauge fields restore the local 
symmetry broken in mean field theory.

However, it is not clear that the slave-boson mean field theories are a
good starting point (except for the cases where a physical phase
transition into a magnetic or superconducting phase takes place). 
Alternatively one may guarantee the local gauge invariance of
approximations by deriving those from a generating functional 
\cite{kadanoff.61}. The
local gauge invariance ensures conservation of the local occupation
at each lattice site (the ``local charge'') in time. In order to
effect the actual projection, it is still necessary to fix the
occupation numbers at a given time locally. So far this has only
been achieved for impurity models, e.g. the Anderson model of a
magnetic impurity in a metallic host \cite{hewson.93,coleman.84,bickers.87}.

Even if the projection is done exactly, an equally important question
is the selection of the dominating contributions in perturbation
theory in the hybridization (or in the hopping integrals and exchange
interaction in the case of lattice models). In this paper we
address this question for the infinite U Anderson impurity model 
(degeneracy $N=2$). We employ the numerical renormalization group (NRG)
to calculate the slave-boson and pseudo-fermion  spectral functions and
compare the results with those of the simplest fully projected
conserving approximation, the so called ``non--crossing approximation''
(NCA) \cite{keiter.nca,kuramoto.nca}. While the NCA is known to give 
excellent results for large
degeneracy $N$ and not too low temperatures, including the cross--over
from low ($T<<T_{K}$) to high temperature ($T>>T_{K}$, where $T_{K}$
is the Kondo temperature)\cite{coleman.84,kuramoto.nca}, 
pronounced deviations from exactly known
results appear at $N=2$\cite{bickers.87,mh.84}. The comparisons will 
allow us to pinpoint the deficiencies of the NCA and to identify 
possible improvements.

In earlier publications \cite{costi.94,costi.94b} we have presented 
results of an NRG calculation of the slave--boson and pseudo--fermion 
spectral functions at zero temperature. 
We found these functions to be infrared divergent
at threshold, with critical exponents dependent on the d-level
occupancy $n_{d}$ and given by simple expressions identical to
the well known X--ray absorption threshold exponents.

The paper is organized as follows: In Section II we formulate the
model, describe how we implement the numerical renormalization group
for studying auxiliary spectral functions of this model and describe
the NCA within the framework of conserving approximations. Section III
describes our results for spectral functions calculated with the above
two methods. In Section IV we summarize our main results.
Some details of the auxiliary particle technique, which make it suitable 
for an effective evaluation at the lowest temperatures, are discussed in the
appendices.

\section{Formulation}
\label{sec:II}

\subsection{Model and Application of the Renormalization Group}
\label{sec:IIa}

The Anderson model of an impurity $d$ electron state hybridizing with 
the conduction band with infinitely strong Coulomb repulsion in the
d--level in auxiliary particle representation takes the form
\begin{equation}
H = H_{c} + \epsilon_{d}\sum_{\sigma}f_{\sigma}^{\dagger}f_{\sigma} +
V\sum_{\sigma}(c_{0\sigma}^{\dagger}b^{\dagger}f_{\sigma} + h.c.),
\label{eq:AM}
\end{equation} 
where
$H_{c}=\sum_{k\sigma}\epsilon_{k}c_{k\sigma}^{\dagger}c_{k\sigma}$ is
the conduction electron kinetic energy and
$c_{0\sigma}=\sum_{k}c_{k\sigma}$ annihilates a conduction electron
with spin $\sigma$ at the impurity site $0$. 
The Hilbert space consists of disjoint subspaces characterized by the
conserved auxiliary particle number 
$Q = b^{\dagger}b+\sum_{\sigma}f_{\sigma}^{\dagger}f_{\sigma}$,
$Q=0,1,2,...$. The physical subspace is defined by the constraint
$Q=1$.  Following Wilson
\cite{nrg.ref} we (i) linearize the spectrum of $H_c$ about the Fermi 
energy $\epsilon_{k}\rightarrow k$, 
(ii) introduce a logarithmic mesh of k points $k_{n}=\Lambda^{-n}$
and (iii) perform a unitary transformation of the $c_{k\sigma}$ such
that $c_{0\sigma}$ is the first operator in the new basis and $H_{c}$
takes the form of a tight--binding Hamiltonian in k--space,
\begin{equation}
H_{c} =
\sum_{n=0}^{\infty}\sum_{\sigma}\xi_{n}\Lambda^{-n/2}
(c_{n+1\sigma}^{\dagger}c_{n\sigma}+ h.c.),\label{eq:tight-binding}
\end{equation}
with $\xi_{n}\rightarrow (1+\Lambda^{-1})/2$ for $n>>1$.
These steps are explained in detail in
\cite{nrg.ref}
and can be taken over for the present model without change.

The Hamiltonian (\ref{eq:AM}) together with the discretized form of the
kinetic energy (\ref{eq:tight-binding}) in the new basis is now 
diagonalized by the following iterative process: (i) one defines 
a sequence of finite size Hamiltonians $H_{N}$ by replacing 
$H_{c}$ in (\ref{eq:tight-binding}) by
$H_{N}^{c} =
\sum_{n=0}^{N-1}\sum_{\sigma}\xi_{n}\Lambda^{-n/2}
(c_{n+1\sigma}^{\dagger}c_{n\sigma}+ h.c.)
\label{eq:tight-binding-truncated};$
(ii) starting from 
$H_{0}=\epsilon_{d}\sum_{\sigma}f_{\sigma}^{\dagger}f_{\sigma} +
V\sum_{\sigma}(c_{0\sigma}^{\dagger}b^{\dagger}f_{\sigma} + h.c.)$, 
each successive hopping may be considered
as a perturbation on the previous Hamiltonian; (iii) the Hamiltonians
$H_{N}$ are scaled such that the energy spacing remains the same.
This defines a renormalization group transformation
$\bar{H}_{N+1} = \Lambda^{1/2}\bar{H}_{N} + \xi_{N}
\sum_{\sigma}(c_{N+1\sigma}^{\dagger}c_{N\sigma}+h.c.) - \bar{E}_{G, N+1}$,
with $\bar{E}_{G,N+1}$ chosen so that the ground state energy
of $\bar{H}_{N+1}$ is zero. The Hamiltonians $\bar{H}_{N}$ are diagonalized
within subspaces of well defined $Q,N_{e},S,S_{z}$ where 
$N_{e}$ is the total fermion number, $S$ the total spin and
$S_{z}$ the z--component of total spin. The use of these conserved quantities
leads to significant reductions in the size of matrices to be diagonalized, 
however the dimension of $\bar{H}_{N}$ 
grows as $4^{N}$ and it is necessary to truncate the higher energy states 
for $N>7$. Approximately $1/4$ of the states generated at each iteration
are retained in the calculations and this constitutes a surprisingly accurate
approximation for the present model. 
This accuracy is evidenced by the fact that
various exact relations, such as the Friedel sum rule, which relates the 
impurity spectral density at the Fermi level to the local level occupancy, 
are satisfied to a high degree of accuracy \cite{costi.94c}. In addition, 
the number of states retained per iteration, $N_{st}$, which is a free 
parameter, can be varied from 250 to 2000 states without any significant 
change in the results presented here. This, together with the almost 
cut--off independent results for $\Lambda \ge 1.5$ indicates the 
convergence and accuracy of the method.

\subsection{NRG Calculation of Auxiliary Spectral Functions}
\label{sec:IIb}

Within the framework of the NRG it is natural to represent the 
Green functions in an eigenbasis of the Hamiltonian. In the
enlarged Hilbert space the retarded propagators for pseudo--fermions
and slave--bosons are defined by
\begin{eqnarray}
G_{f\sigma}(\omega+i0,T,\lambda) & = & 
\frac{1}{Z_{GC}}\sum_{Q,m,n}|M_{m,n}^{f}|^{2}
(e^{-\beta(\epsilon_{n} + \lambda Q)}
+ e^{-\beta(\epsilon_{m}+\lambda (Q+1))})
/(\omega+i0-\lambda-(\epsilon_{m}-\epsilon_{n})),\nonumber\\
G_{b}(\omega+i0,T,\lambda) & = & \frac{1}{Z_{GC}}\sum_{Q,m,n}|M_{m,n}^{b}|^{2}
(e^{-\beta(\epsilon_{n} + \lambda Q)}
- e^{-\beta(\epsilon_{m}+\lambda (Q+1))})
/(\omega+i0-\lambda-(\epsilon_{m}-\epsilon_{n})),\nonumber
\end{eqnarray}
where, for each $Q$, $\epsilon_m, \epsilon_n$ 
are eigenvalues of the subspaces $Q+1,Q$, 
where $Z_{GC}(T,\lambda)$ is the grand--canonical partition function,
\begin{equation}
Z_{GC}(T,\lambda) = \sum_{Q,m}e^{-\beta (\epsilon_{m}+\lambda Q)} 
= Z_{Q=0}+e^{-\beta\lambda} Z_{Q=1}+\dots\label{eq:gc-partition}
\end{equation}
and $M_{m,n}^{O}=<Q+1,m|O^{\dagger}|Q,n>$ with $O=f_{\sigma},b$ are the
many--body matrix elements for pseudo--fermions and slave--bosons,
respectively. The constrained propagators are then obtained as
\begin{eqnarray}
G_{f,b}(\omega+i0,T,\lambda\rightarrow\infty) & = & 
\frac{1}{Z_{Q=0}}\sum_{m,n}|M_{m,n}^{f,b}|^{2}e^{-\beta\epsilon_{n}}
/(\omega+i0-(\epsilon_{m}-\epsilon_{n})),
\end{eqnarray}
where the frequency has been shifted as
$\omega\rightarrow\omega+\lambda$. We are interested in the
following projected spectral functions (see Appendix~C for details))
$A_{f,b}^{+}(\omega,T)=-\lim_{\lambda\rightarrow\infty}[\mbox{Im}\;
{G_{f,b}(\omega,T,\lambda)}/\pi]$ and 
$A_{f,b}^{-}(\omega,T)=-\lim_{\lambda\rightarrow\infty}
\frac{e^{-\beta\omega}}{Z_{C}(T)}
[\mbox{Im}\;{G_{f,b}(\omega,T,\lambda)}/\pi]$,
\begin{eqnarray}
A_{f,b}^{+}(\omega,T) & = & 
\frac{1}{Z_{Q=0}(T)}
\sum_{m,n}|<1,m|O^{\dagger}|0,n>|^{2}e^{-\beta\epsilon_{0,n}}
\delta(\omega-(\epsilon_{1,m}-\epsilon_{0,n})),\label{eq:aplus}\\
A_{f,b}^{-}(\omega,T) & = & 
\frac{1}{Z_{C}(T)Z_{Q=0}(T)}\sum_{m,n}|<1,m|O^{\dagger}|0,n>|^{2}
e^{-\beta\epsilon_{1,m}}
\delta(\omega-(\epsilon_{1,m}-\epsilon_{0,n})),\label{eq:aminus}
\end{eqnarray}
In the definition of $A^-_{f,b}$ $Z_C=Z_{Q=1}$ is introduced 
in order to obtain a well defined zero temperature limit\footnote{
The functions $A_{f}^{+}$,
$A_{b}^{+}$ defined here correspond to the $A$, $B$ functions, respectively 
in Ref.~\cite{coleman.84}, and similarly the functions 
$A_{f}^{-}$, $A_{b}^{-}$ correspond to the $a$ and $b$ functions.}.
At zero temperature the spectral functions reduce to
\begin{eqnarray}
A_{f,b}^{+}(\omega,T=0) & = & \frac{1}{Z_{Q=0}(0)}
\sum_{m}|<1,m|O^{\dagger}|\Phi_{0}>|^{2}
\delta(\omega+E^{GS}_{Q=0}-\epsilon_{1,m}),\label{eq:aplus0}\\
A_{f,b}^{-}(\omega,T=0) & = &  \frac{1}{Z_{C}(0)\; Z_{Q=0}(0)}
\sum_{n}|<\Phi_{1}|O^{\dagger}|0,n>|^{2}
\delta(\omega+\epsilon_{0,n}-E^{GS}_{Q=1}).\label{eq:aminus0}
\end{eqnarray}
Here $|\Phi_{0}>$ 
is the groundstate of the $Q=0$ subspace
of non--interacting conduction electrons and $|1,m>$ are the excited states
of the $Q=1$ subspace of the interacting system, $E^{GS}_{Q=0}$
and $\epsilon_{1,m}$ are the corresponding energy eigenvalues. The spectral
functions $A_{f,b}^{+}(T=0)$ vanish identically {\em below} the
threshold $E_{0}=E^{GS}_{Q=1}-E^{GS}_{Q=0}$. 
Similarly, in the expression for $A_{f,b}^{-}(T=0)$, 
$|\Phi_{1}>$ is the groundstate of the interacting system 
($Q=1$ subspace) and $|0,n>$ are the excited states
of the non--interacting conduction electron system, $E^{GS}_{Q=1}$ and
$\epsilon_{0,n}$ the corresponding energy eigenvalues. These
spectral functions vanish {\em above} the threshold energy $E_{0}$.
From (\ref{eq:sum-rule1}-\ref{eq:sum-rule2}) and (\ref{eq:aplus0}) we see 
that the $T=0$ spectral functions $A_{f,b}^{+}$ satisfy the sum rule
\begin{equation}
\int_{-\infty}^{+\infty}A_{f,b}^{+}(\omega,0)d\omega = 1
\end{equation}

In practice, within the NRG technique, it is the excitation energies 
from the respective $Q=0$ or $Q=1$ groundstates which are calculated. 
Hence, it is convenient to set the threshold energy, $E_{0}$, to zero. 
The $T=0$ projected auxiliary spectral functions then have divergences 
at zero energy. As described in Appendix~D it is also possible to
formulate the NCA equations of section C such that the NCA spectral functions
exhibit their $T=0$ divergences at zero energy. This makes it easier to
compare with the NRG results and in addition improves the accuracy of
the NCA solution at the lowest temperatures. The reference energy scale 
for the physical $d$ electron Green's function,
$G_{d\sigma}(\omega,T)=<<b^{\dag}f_{\sigma};b f_{\sigma}^{\dag}>>_{Q=1}$,
\begin{eqnarray}
G_{d\sigma}(\omega+i0,T) & = & 
\frac{1}{Z_{C}}\sum_{m,n}|M_{m,n}^{d}|^{2}
(e^{-\beta \epsilon_{n}}
+ e^{-\beta \epsilon_{m}})
/(\omega+i0-(\epsilon_{m}-\epsilon_{n})),
\end{eqnarray}
is independent of $E_0$ and is set by the Fermi level $\epsilon_F=0$. 

The matrix elements $M_{m,n}^{f,b}$ of the pseudo-particle operators 
$f_{\sigma}^{\dagger}$, $b^{\dagger}$ in
(\ref{eq:aplus}--\ref{eq:aminus}) are calculated recursively using the
formulae given in Appendix~A. Similar formulae apply to the matrix elements
$M_{m,n}^{d}$ for the physical $d$ electron Green's function. 
For each iteration step, they are substituted together with 
the energy eigenvalues into (\ref{eq:aplus0}--\ref{eq:aminus0}) 
to give the $T=0$ spectral functions $\bar{A}_{N,f,b}^{\pm}(\omega)$. 
In principle if all states up to stage $N$ were retained, $H_{N}$
would describe excitations on all energy
scales from the band edge $D=1$ down to the lowest energy scale
present in $H_{N}$, i.e. $\omega_{N}$. Due to the elimination of
higher energy states at each step, the actual range of excitations in
$H_{N}$ is restricted to $\omega_{N}\le \omega \le K\omega_{N}$, where
$K \approx 7$ for $\Lambda \approx 2$ retaining $500$ states
per iteration. Thus at step $N$, the spectral functions are calculated at an
excitation energy $\omega \approx 2\omega_{N}$ in the above range. 
The delta functions in (\ref{eq:aplus0}--\ref{eq:aminus0}) are 
broadened with Gaussians of width $\alpha_{N} \approx \omega_{N}$ 
appropriate to the energy level structure of $H_{N}$ \cite{costi.94c}.

Finally we note that the constraint of no double occupancy, $Q=1$, is
implemented exactly within the NRG calculations since $Q$ is a conserved
quantum number. Appendix~C describes in detail the implementation of
this constraint for the conserving approximation of following section.

\subsection{Conserving approximation: NCA}
\label{sec:IId}

For small hybridization $V$ compared to the impurity level
$\epsilon _d$, a perturbation expansion in terms of $V$ seems reasonable.
It is useful to express the auxiliary particle Green's functions for
given ``chemical potential'' $-\lambda$ defined in (\ref{eq:projection}) 
in terms of self--energies $\Sigma_{b}(\omega,T,\lambda)$ and
$\Sigma_{f}(\omega,T,\lambda)$ as
\begin{eqnarray}
G_{f\sigma}(\omega,T,\lambda) & = & [\omega - \lambda - \epsilon _{d} -
\Sigma_{f}(\omega,T,\lambda)]^{-1}\\
G_{b}(\omega,T,\lambda) & = & [\omega - \lambda - 
\Sigma_{b}(\omega,T,\lambda)]^{-1}
\end{eqnarray}
where $\omega$ takes the values of fermionic or bosonic Matsubara frequencies.
The projected spectral functions are obtained by shifting the frequency 
$\omega \rightarrow \omega + \lambda $ and taking the limit 
$\lambda\rightarrow\infty$, as described in Appendix~C. 
The local Green's function of conduction 
electrons $G_{c\sigma}(\omega,T,\lambda)$ is given likewise in terms
of the self energy $\Sigma_{c\sigma}(\omega,T,\lambda)$ as
\begin{equation}
G_{c\sigma}(\omega,T,\lambda) = [(G_{c\sigma}^{0})^{-1} - 
\Sigma_{c\sigma}(\omega,T,\lambda)]^{-1}
\end{equation}
where 
$G_{c\sigma}^{0}(\omega) = \int d\epsilon N(\epsilon)/(\omega - \epsilon)$
is the local Green's function of the bare conduction band, with $N(\epsilon)$
the density of states. 
There is an exact relation between
the d-electron Green's function and $\Sigma_{c}$:
\begin{equation}
G_{d\sigma}(\omega,T,\lambda) = \frac{1}{V^{2}}
\frac{\Sigma_{c\sigma}(\omega,T,\lambda)}
{1-\Sigma_{c\sigma}(\omega,T,\lambda)\; G_{c\sigma}^{0}(\omega )}.
\label{eq:Gd2}
\end{equation}
Thus, the principal problem remaining in the evaluation of the theory 
is the proper choice of approximation in which the three self-energies
are calculated. The slave--boson Hamiltonian (\ref{eq:AM}) is invariant
under two independent $U(1)$ gauge transformations: (1) simultaneous
transformation of the pseudo--fermion and the slave--boson operators
and (2) simultaneous transformation of the pseudo--fermion and the 
conduction electron operators. These symmetries correspond to the 
conservation of the auxiliary particle number and the total number 
of fermions, respectively:
\begin{eqnarray}
\sum _{\sigma} f^{\dag}_{\sigma }f_{\sigma } + b^{\dag}b & = & const. 
\label{eq:naux}\\
\sum_{\sigma}f^{\dag}_{\sigma }f_{\sigma } + 
\sum_{k\sigma} c^{\dag }_{k\sigma}c_{k\sigma} & = & const.
\label{eq:nfermi}
\end{eqnarray}
The Green's functions $G_{f}$, $G_{b}$ are not invariant under local
(in time) gauge transformations, but we must require $G_{d}$ to be invariant.
Constructing a gauge invariant approximation is, a priori, a non--trivial
task, since the $d$ electron number $n_{d}$ is not a
conserved quantity. Yet, it may be shown that a gauge invariant
approximation obeying both conservation laws (\ref{eq:naux}), 
(\ref{eq:nfermi}) is constructed by deriving all three self--energies from
one generating functional $\Phi$ :
$\Sigma_{f\sigma}=\delta\Phi/\delta{G_{f\sigma}}$, 
$\Sigma_{b}=\delta\Phi/\delta{G_{b}}$,
$\Sigma_{c\sigma}=\delta\Phi/\delta{G_{c\sigma}}$. 
Eq.~(\ref{eq:Gd2}) then provides the rule for a gauge invariant 
approximation for $G_d$. The functional $\Phi$
is given in terms of closed skeleton diagrams with suitable combinatorial
factors. The lowest order contribution to $\Phi$ is of second order in
$V$, $\Phi = - V^{2}\frac{1}{\beta}\sum_{\omega}\sum_{\epsilon}
G_{f\sigma}(\omega)G_{c\sigma}(\epsilon)G_{b}(\omega-\epsilon)$ (Fig.~1 a).
Functional differentiation yields the self--energies
\begin{eqnarray}
\Sigma_{f\sigma}(\omega) & = & -V^{2}\frac{1}{\beta}
\sum_{\epsilon}G_{c\sigma}(\epsilon)
G_{b}(\omega-\epsilon)\\ 
\Sigma_{b}(\Omega) & = & V^{2}\frac{1}{\beta}
\sum_{\epsilon}G_{f\sigma}(\Omega+\epsilon)
G_{c\sigma}(\epsilon)\\
\Sigma_{c\sigma}(\omega) & = & -V^{2}\frac{1}{\beta}
\sum_{\epsilon}G_{f\sigma}(\epsilon)
G_{b}(\epsilon-\omega).
\end{eqnarray}
After the transformation $\omega \rightarrow \omega + \lambda$, 
$\lambda\rightarrow\infty$, one finds explicitly
\begin{eqnarray}
\Sigma_{f\sigma}(\omega,T) & = & V^{2}\sum_{k}(1-f(\epsilon_{k}))
G_{b}(\omega-\epsilon_{k})\label{eq:nca-selff}\\
\Sigma_{b}(\omega,T) & = & V^{2}\sum_{k,\sigma}f(\epsilon_{k})
G_{f\sigma}(\omega+\epsilon_{k}),\label{eq:nca-selfb}\\
\Sigma_{c\sigma}(\omega,T,\lambda\rightarrow\infty) 
& = & V^{2}e^{-\beta\lambda}\int \mbox{d}\epsilon
e^{-\beta\epsilon}\bigl[ G_{f\sigma}(\epsilon +\omega)A_b^+(\epsilon)-
A_f^+(\epsilon )G_b(\epsilon -\omega )\bigr] 
\label{eq:nca-selfc}
\end{eqnarray}
and $G_{f\sigma}(\omega)^{-1}=\omega-\epsilon_{d}-\Sigma_{f\sigma}(\omega)$,
$G_{b}(\omega)^{-1}=\omega-\Sigma_{b}(\omega)$,
where $f(\epsilon_{k})$ is the Fermi function. The physical d-electron
Green's function is obtained from the limiting procedure 
$G_{d\sigma}(\omega,T)=
\lim_{\lambda\rightarrow\infty}e^{\beta\lambda}G_{d\sigma}(\omega,T,\lambda)$
(compare Appendix~C).
As $\Sigma_{c\sigma}(\omega,T,\lambda)\sim e^{-\beta\lambda}$ 
for $\lambda\rightarrow\infty$, the self--energy corrections to $G_{c}$
and to the denominator of $G_{d}$ vanish, and
\begin{equation}
G_{d\sigma}(\omega,T)=\frac{1}{V^{2}}
\lim_{\lambda\rightarrow\infty}e^{\beta\lambda}
\Sigma_{c\sigma}(\omega,T,\lambda).
\end{equation}
The impurity spectral density follows as,
\begin{eqnarray}
\rho_{d}^{NCA}(\omega,T) & = & -\frac{1}{\pi}
Im\;\;G_{d\sigma}(\omega+i0,T)\nonumber\\
& = & \int_{-\infty}^{+\infty}d\epsilon
[A_{f}^{+}(\epsilon+\omega,T)A_{b}^{-}(\epsilon,T)
+A_{f}^{-}(\epsilon,T)A_{b}^{+}(\epsilon-\omega,T)],
\label{eq:nca-convolution}
\end{eqnarray}
which at $T=0$ reduces to,
\begin{eqnarray}
\rho_{d}^{NCA}(\omega,T=0) & = & 
\Theta(\omega)\int_{E_{0}-\omega}^{E_{0}}d\epsilon
A_{f}^{+}(\epsilon+\omega,T=0)A_{b}^{-}(\epsilon,T=0)\nonumber\\
& + & \Theta(-\omega)\int_{E_{0}+\omega}^{E_{0}}d\epsilon 
A_{f}^{-}(\epsilon,T=0)A_{b}^{+}(\epsilon-\omega,T=0).
\end{eqnarray}
The above approximation is known as the ``non--crossing approximation'' 
(NCA) because it does not include any diagrammatical contributions with
crossed conduction electron lines \cite{kuramoto.nca}.

In general the impurity electron Green's function may be expressed 
with the help of a vertex function $\Lambda (\epsilon,\omega)$ as (Fig.~1 b)
\begin{equation}
G_{d\sigma}(\omega) = -\lim_{\lambda\rightarrow\infty}e^{\beta\lambda}
\sum_{\epsilon}G_{f\sigma}(\epsilon)G_{b}(\epsilon-\omega)
\Lambda (\epsilon,\omega)\label{eq:realgf+vertex}
\end{equation}
As shown below, there is reason to expect that the vertex function $\Lambda$
plays an important role.

\section{Results}
\label{sec:III}

The NRG calculations were performed for $\Lambda=2$,
keeping $250$ states per iteration for each subspace ($Q=0,1$).
The hybridization strength $\Delta=\pi\;V^{2}\rho(\epsilon_{F})=\pi\;V^{2}/2D$
was chosen to be $0.01D$ with the half--bandwidth $D=1$. 
Several values of the local
level position $\epsilon_{d}$ were chosen in order to characterize the
behaviour of the spectral densities in the various regimes. The
NRG spectral functions were evaluated at $T=0$ and the NCA spectral
functions were evaluated following appendix D for the same set of
parameters and for temperatures down to $T=10^{-6}D << T_{0}$,
where $T_{0}$ is the low energy scale of the model. 
This was sufficiently low to allow comparison with the $T=0$ NRG 
results over most of the interesting energy range. We define
$T_0$ to be the Kondo temperature,
\begin{equation}
k_{B}T_{K} = D\sqrt{\frac{\Delta}{D}} e^{-\pi\epsilon_{d}/2\Delta}
\label{eq:kondo-temp}
\end{equation}
in the Kondo regime $\epsilon_{d}/\Delta \leq -2$, $\Delta$ in the mixed valent
regime $|\epsilon_{d}/\Delta| \leq 1$ and $\epsilon_{d}$ in the empty orbital
regime $\epsilon_{d}/\Delta >> 1$.

\subsection{Threshold behaviour of the NRG auxiliary spectral functions}
\label{sec:IIIa}

The $T=0$ auxiliary spectral functions diverge
at the threshold $E_{0}$ as shown in Fig.\ (\ref{sd-kondo}--\ref{sd-empty}). 
This behaviour may be understood as a result of the orthogonality 
catastrophe theorem \cite{anderson.67}. To see this more clearly 
we re--formulate the spectral densities in 
(\ref{eq:aplus0}--\ref{eq:aminus0}) in the following way,
\begin{eqnarray}
A_{f,b}^{+}(\omega,T=0) & = 
&\frac{1}{Z_{Q=0}}\sum_{m}|<1,m|\tilde{\Phi}_{0}>|^{2}
\delta(\omega-\epsilon_{1,m}),\label{eq:over1}\\
A_{f,b}^{-}(\omega,T=0) & = 
& \frac{1}{Z_{Q=0}}\sum_{n}|<\tilde{\Phi}_{1}|0,n>|^{2}
\delta(\omega+\epsilon_{0,n})\label{eq:over2}.
\end{eqnarray}
In the above, $|\tilde{\Phi}_{0}>=O^{\dagger}|{\Phi}_{0}>$,
with $O=f_{\sigma},b$, represents the non--interacting ($U=0$) 
groundstate with 1 ($O=f^{\dag}_{\sigma}$) or 0 ($O=b^{\dag}$) 
local electrons present. Similarly, $|\tilde{\Phi}_{1}>=O|\Phi_{1}>$, 
with $O=f_{\sigma},b$, represents the interacting ($U=\infty$) 
groundstate with 1 ($O=f^{\dag}_{\sigma}$) or 0 ($O=b^{\dag}$) 
local electrons present. In this formulation, we see that 
$A_{f,b}^{+}$ measures the overlap density between the groundstate 
of the non--interacting ($U=0$)  
band electrons with 0 or 1 local electrons present with the excited 
states of the interacting $U=\infty$ Hamiltonian. Similarly, $A_{f,b}^{-}$
measures the overlap density between the groundstate of the 
interacting $U=\infty$ Hamiltonian with 0 or 1 local electrons present and 
the excited states of the non--interacting band electrons. 
This interpretation is identical to that for the core level
spectral functions in the X--ray problem. The
analogy is useful but requires care since the matrix elements
in (\ref{eq:aplus}--\ref{eq:aminus}) are no longer between
two non--interacting systems as in the X--ray problem.
This leads in particular to a new energy scale, $T_{0}$, for the onset of
the asymptotic power law behaviour, which is $T_{K}$, $\Delta$ or
$\epsilon_{d}$ in the Kondo, mixed valent and empty orbital regimes,
respectively. We find that it is only in the Fermi liquid regime,
$|\omega-E_{0}| << T_{0}$, that the power law behaviour is well
characterized. The approach to this asymptotic power law
is faster for the boson spectral functions than for the fermion spectral
functions in all cases. We note that within the NCA, the approach to
the threshold behaviour with $n_{d}$ independent exponents $\alpha_{f}=1/3$
and $\alpha_{b}=2/3$ is also only asymptotic and requires in particular
going down to temperatures $T < 10^{-2}T_{0}$ in order to see these
exponents.

The threshold exponents for the slave--boson and pseudo--fermion spectral
functions were extracted by numerically differentiating the spectral
functions. Typically, well defined exponents can be extracted only for
energy scales $|\omega-E_{0}| < 10^{-2} T_{0}$.
The exponents are shown in Fig.\ (\ref{exponents}) and Table~\ref{table1}
as a function of $n_{d}$, the local level occupancy at $T=0$. 
The latter, shown in Fig.\ (\ref{valence}), was calculated 
by evaluating $n_{d}(T)$ from the partition
function at a sequence of decreasing temperatures $T_{N}\sim
\Lambda^{-(N-1)/2}$ and then taking the limit $T\rightarrow 0$. 

Remarkably, the threshold exponents turn out to be the usual photoemission 
and absorption exponents for the X--ray problem and are given in terms
of the conduction electron phase--shift at the Fermi level \cite{costi.94},
$\delta_{\sigma}=\delta_{\sigma}(\epsilon_{F})$, by
\begin{eqnarray}
\alpha_{f} & = & n_{d} - \frac{n_{d}^{2}}{2} = 2\frac{\delta_{\sigma}}{\pi} -
\sum_{\sigma}(\frac{\delta_{\sigma}}{\pi})^{2}\label{eq:fermion-exp}\\
\alpha_{b} & = & 1 - \frac{n_{d}^{2}}{2} =
1 - \sum_{\sigma}(\frac{\delta_{\sigma}}{\pi})^{2}\label{eq:boson-exp} 
\end{eqnarray}
where the last equations on the RHS of 
(\ref{eq:fermion-exp}--\ref{eq:boson-exp}) follow from the
Friedel sum rule, $\delta_{\sigma}=\pi\;n_{d}/2$.
These results are clearly illustrated in Fig.\ (\ref{exponents}) where the
functions $n_{d} - n_{d}^{2}/2$ and $1 - n_{d}^{2}/2$ are plotted
against $n_{d}$ together with the exponents $\alpha_{f,b}$ deduced 
from the spectral functions. The exponents $\alpha_{f,b}$ agree with
the RHS of (\ref{eq:fermion-exp}--\ref{eq:boson-exp}) to 3 significant
figures in nearly all cases and are the same below and above the threshold, 
\begin{equation}
A_{f,b}^{\pm} = a_{f,b}^{\pm}\;|\omega - E_{0}|^{-\alpha_{f,b}}
\label{eq:asymptotic}
\end{equation}
A qualitative argument based on charge neutrality considerations has
been given for the above form of the exponents \cite{costi.94}.
We note that the same functional form of the exponents on the phase
shift (\ref{eq:fermion-exp}--\ref{eq:boson-exp}) is also found in the 
spinless model with constraint \cite{costi.94b} in agreement with exact 
analytic results \cite{mh.88}. The single phase shift 
in this case is given by $\delta=\pi n_{d}$ . An $n_{d}$ dependent 
exponent was also found by Read in considering how Gaussian
and higher order corrections restore the gauge symmetry broken by the
slave--boson mean field theory \cite{read.85} . 
Generalizing the above threshold exponents to the N--fold
degenerate model we have 
$\alpha_{f}=2\delta_{m}/\pi -\sum_{m}(\delta_{m}/\pi)^{2}
=2n_{d}/N - n_{d}^{2}/N$ and
$\alpha_{b}=1 -\sum_{m}(\delta_{m}/\pi)^{2}
=1-n_{d}^{2}/N$ where $m$ labels the scattering channels. 
The same exponents were found for the $N$--fold degenerate Anderson 
impurity model in perturbative calculations to order $V^{4}$ \cite{mh.88}
adding to the plausibility of the above generalization. The above conjecture
for $\alpha_{f,b}$ is in disagreement with recent results obtained in 
the limit $n_{d}\rightarrow 1$ and in the large $N$ expansion including
order $1/N^{2}$ \cite{gruneberg.91}, 
$\tilde{\alpha}_{f}=\frac{1-\frac{1}{N^{2}}}{N+1-\frac{1}{N^2}}$, 
$\tilde{\alpha}_{b}=\frac{N-\frac{2}{N^2}}{N+1-\frac{1}{N^2}}$. These
results were obtained using a perturbative renormalization group technique,
which we do not expect to be as accurate as the non--perturbative numerical
scheme used here. We see from our results, 
generalized to arbitrary $N$, that the 
NCA exponents $1/(N+1)=1/N + O(1/N^{2})$ and $N/(N+1)=
1-1/N +O(1/N^{2})$ are correct only in the
limit $n_{d}\rightarrow 1$ and $N\rightarrow\infty$ (or in the trivial
limit $n_{d}\rightarrow 0$). Away from this limit, 
vertex corrections in the auxiliary Green's functions, absent
in the NCA, are therefore important in determining the correct
threshold exponents. The expressions for
the threshold exponents of auxiliary particle propagators in terms of
X--ray photoemission exponents appears to be a general property of several
impurity models exhibiting Fermi liquid fixed points.
\subsection{NRG auxiliary spectral functions at higher energies}
\label{sec:IIIb}

At higher frequencies the following features are observed in the $T=0$ 
spectral functions. In the Kondo regime, Fig. \ (\ref{sd-kondo}), 
there is a peak in the slave--boson
spectral function $A_{b}^{+}$ at $\omega=|\epsilon_{d}|$ and a much less
pronounced feature in the corresponding pseudo--fermion spectral function 
$A_{f}^{+}$.
As $\epsilon_{d}$ is raised through the Fermi level from below 
the peak in $A_{b}^{+}$ at $\omega=|\epsilon_{d}|$ becomes less pronounced
and almost disappears in the mixed valent $\epsilon_{d}/\Delta \sim 0$, 
Fig. \ (\ref{sd-mixed-valent}), and empty orbital regimes 
$\epsilon_{d}/\Delta > 1$ Fig. \ (\ref{sd-empty}). 
In addition, its position is renormalized above the bare 
value $|\epsilon_{d}|$.
At the same time the small feature at positive energies in the 
pseudo--fermion spectral function $A_{f}^{+}$ develops into a well defined 
peak in the mixed valent and empty orbital regimes.
The ``$-$'' spectral functions $A_{f,b}^{-}(\omega,T=0)$  exhibit monotonic
behaviour for all parameter regimes.

\subsection{Comparison of NRG and NCA auxiliary spectral functions}
\label{sec:IIIc}

In comparing NRG and NCA spectral functions, three energy regimes
should be distinguished:
\begin{description}
\item[(I)] asymptotically low energy regime, $|\omega-E_{0}|/T_{0} << 1$,
\item[(II)] crossover regime, $|\omega-E_{0}|/T_{0} \sim 1$,
\item[(III)] high energy regime, $|\omega-E_{0}|/T_{0} >> 1$.
\end{description}
The energy range (I) corresponds to that discussed in the
section on threshold exponents. Here we discuss the energy ranges between
(I) and (II), and between (II) and (III). In 
Fig. \ (\ref{sd-kondo-nca}--\ref{sd-empty-nca}) 
the same qualitative trends described in the previous section for the
NRG auxiliary spectral functions can be seen in 
the corresponding NCA solutions. The NCA results for a finite but
very low temperature $T=10^{-6}D$ are compared to the corresponding
$T=0$ NRG results. Surprisingly good quantitative agreement is
seen in the slave--boson spectral function $A_{b}^{+}$ above the 
threshold down to energy scales well below $T_{0}$ in all regimes.
The agreement is particularly good in the Kondo regime for 
$-4\le\epsilon_{d}/\Delta\le-2$, where it extends down to $10^{-2}T_{0}$
(e.g., Fig. \ref{sd-kondo-nca}a).
The spectral function $A_{b}^{-}$ below the threshold also shows 
good agreement with the NRG result in the Kondo regime with decreasing
agreement in the mixed valent $\epsilon_{d}/\Delta \sim 0$ and 
empty orbital regimes $\epsilon_{d}/\Delta >> 1$. 
Turning now to the pseudo--fermion spectral functions
we see that there is again good quantitative agreement between NCA and
NRG for $A_{f}^{+}$ above the threshold and for all energy scales down to
$T_{0}$. This is true in all regimes. Below the threshold the agreement for
the $A_{f}^{-}$ spectral function even extends to well below $T_{0}$, except
in the Kondo regime for $\epsilon_{d}/\Delta \le -4$ where we could not
obtain quantitative agreement except in the region 
$10^{-1}\le\omega\le 10^{1}$.
From these comparisons we see that the most serious differences, as far as
low energy behaviour is concerned, between
the NCA and NRG auxiliary spectral functions are in $A_{f}^{+}$ for 
$\omega < T_{0}$ in the Kondo case and in $A_{b}^{-}$ in the mixed 
valent and empty orbital cases. 
The latter we attribute to the inaccuracy of the NCA in the energy range 
between (I) and (II) in the mixed valent and empty orbital regimes 
to be described further in the section on impurity spectral densities.  
A more interesting discrepancy which arises from these comparisons is the
former. The functions $A_{f,b}^{+}$ in NCA are related by self--consistency 
equations (\ref{eq:nca-selff}--\ref{eq:nca-selfb}) derived by second order
perturbation theory in $V$. As pointed out in \cite{kroha.tmatrix}, coherent
spin flip processes considered to be responsible for the Kondo resonance
are not included in the NCA. There are reasons to expect that the 
self--consistent T-matrix approximation proposed in \cite{kroha.tmatrix}
will capture the essential contributions. Recent calculations within the
conserving self--consistent T--matrix approximation indicate that such
improvements do indeed arise \cite{kroha.privatecommun.95}.

A different extension of NCA, called ``post--NCA'' has been proposed
recently on the basis of a $\frac{1}{N}$ expansion of the N-orbital 
model \cite{anders.vertex}. It represents a self--consistent scheme
including vertex renormalizations, which is exact to order $1/N^{2}$.

\subsection{Comparison of NRG and NCA impurity spectral functions}
\label{sec:IIId}

In the previous section we noted that the NCA auxiliary spectral functions
were surprisingly close to the NRG ones for energies down to at least 
$\omega=T_{0}$ and typically they were even quantitatively accurate down
to $\omega<<T_{0}$. Improvements are primarily important in
the auxiliary spectral functions in two areas, to restore the correct
behaviour of the pseudo--fermion spectral function $A_{f}^{+}$ below $T_{0}$
and to recover the exact threshold exponents given by the NRG. 
We now turn to the comparisons for the impurity spectral function and
discuss the role of vertex corrections on the dynamics of the physical
electrons.
The impurity spectral densities are shown in
Fig. \ \ (\ref{rsd-kondo}--\ref{rsd-empty}) where in addition to the NCA, 
$\rho_{d}^{NCA}(\omega)$, and NRG spectral functions, $\rho_{d}^{NRG}(\omega)$,
we also show the impurity spectral function, $\rho_{d}^{NRG-}$, 
obtained by convoluting the NRG auxiliary spectral functions as in 
(\ref{eq:realgf+vertex}) but without the vertex part, i.e.
\begin{equation}
\rho_{d}^{NRG-}(\omega,T) = \int_{-\infty}^{+\infty}d\epsilon[
A_{f}^{+}(\epsilon+\omega,T)A_{b}^{-}(\epsilon,T)
+A_{f}^{-}(\epsilon,T)A_{b}^{+}(\epsilon-\omega,T)]. \label{eq:nrg-convolution}
\end{equation}
In the Kondo regime (Fig.~\ref{rsd-kondo} a-c),
the impurity spectral density shows two peaks,
a charge fluctuation peak close to the local level position $\epsilon_{d}$, 
and a many--body Kondo resonance at the Fermi level. The charge fluctuation
peak is broader in the NRG case, a result of using a logarithmic
discretization for the conduction band which leads to lower resolution 
at higher energies. This is not a fundamental problem with the NRG, and
the resolution of the method at higher energies can be improved by
reducing the discretization parameter $\Lambda$ for the relevant iterations
covering the high energy scales. We have explicitly checked that 
the width and height of the many--body resonance 
at the Fermi level, where the NRG gives the highest resolution, are unaffected
by the broadening used to smooth the delta functions in the discrete
spectra (unless the broadening is made too small which will result 
in uneven spectra). From Fig.\ \ref{rsd-kondo}b-c we see that 
the NCA gives a Kondo resonance which is too broad and too high 
on the BIS ($\omega>0$) side for energies below $\sim 5T_{0}$. On the 
PES side ($\omega<0$), i.e. for $\omega <0$, the agreement with the 
NRG is better. 

The disagreement between the NCA and NRG impurity spectral function for 
energies below $5T_{0}$ is due primarily to the absence of the 
vertex part in the convolution formula for the NCA impurity spectral density 
(\ref{eq:nca-convolution}). This is seen from the good agreement between
$\rho_{d}^{NRG-}$ and $\rho_{d}^{NCA}$ in the range 
$T_{0}\le \omega\le 5T_{0}$ which indicates that the NCA auxiliary
spectral functions are sufficiently accurate in this range and that 
{\em therefore
the difference between $\rho_{d}^{NRG}$ and $\rho_{d}^{NCA}$ must be mainly due
to the neglect of the vertex part in (\ref{eq:nca-convolution})}. 
Similar comparisons for the other Kondo cases $\epsilon_{d}/\Delta=-5,-3$
support the same conclusion. In Table~\ref{table1} we also list 
$\rho_{d}^{NRG}(\omega=0)$ and $\rho_{d}^{NCA}(\omega=0,T=10^{-6}D)$ 
together with the respective relative deviations from the exact 
Friedel sum rule result,
$\rho_{d\sigma}(\omega=0)=\sin^{2}(\pi n_{d}/2)/\pi\Delta$. The NCA
result in the Kondo regime appears reasonable because the singular
behaviour of the impurity spectral density at $T=0$ \cite{mh.84} is
removed by our small finite temperature $T=10^{-6}D$. The exact NCA
result for $\rho_{d}^{NCA}(\omega=0,T=0)$ is actually much worse \cite{mh.84}.

We also show results for the mixed valent, 
Fig. \ (\ref{rsd-mixed-valent}), and empty orbital regimes, 
Fig. \ (\ref{rsd-empty}). There is good agreement between
$\rho_{d}^{NRG}$, $\rho_{d}^{NRG-}$ and $\rho_{d}^{NCA}$ for the
high energy parts of the renormalized resonant level $\omega \ge T_{0}$, but
the incorrect low energy behaviour of the NCA result for $A_{b}^{+}$ 
in the above regimes and the neglect of the vertex part in 
(\ref{eq:nca-convolution}) and 
(\ref{eq:nrg-convolution}) makes both $\rho_{d}^{NCA}$ and 
$\rho_{d}^{NRG-}$ deviate from the exact NRG result at low energies.
The resonant level is approximately a Lorentzian of width $\Delta$, 
and the small asymmetric broadening in the NRG curves is due to 
the logarithmic discretization. An improved description of high energies
could be obtained in both NRG and NCA, if required. Within NRG it is
possible to focus on high energies explicitly by using a smaller 
discretization parameter, $\Lambda$, for the first few iterations. Within
NCA higher energies are easily resolved by using a finer grid to solve the
integral equations at these energies.

Having discussed the effect of vertex corrections on the impurity spectral
function at intermediate $\omega ~\sim T_{0}$ and higher energies 
$\omega > T_{0}$ in the different parameter regimes, we now discuss the
limit $\omega << T_{0}$. In this limit, evaluating the impurity spectral
density, without vertex corrections, i.e. for $\Lambda=1$, using 
(\ref{eq:fermion-exp}--\ref{eq:asymptotic}) in
(\ref{eq:realgf+vertex}) gives
\begin{equation}
\rho_{d\sigma}'(\omega\rightarrow 0^{+}) =
a^{+}_{f}a^{-}_{b}\omega^{1-\alpha_{f}-\alpha_{b}}
B(1-\alpha_{f},1+\alpha_{b})\sim 
|\omega|^{-n_{d}(1-n_{d})}.\label{eq:singularity}
\end{equation}
where $B$ is the Beta function. The exact result at $\omega =0 $ is
given by the Friedel sum rule,
$\rho_{d\sigma}(\omega=0)=\sin^{2}(\pi n_{d}/2)/\pi\Delta$, so 
we conclude that the vertex corrections neglected in (\ref{eq:singularity}) 
are singular at low energies, i.e. close to the threshold, and lead to a 
singularity in $\rho_{d\sigma}$ at the Fermi level 
which cancels that in (\ref{eq:singularity}). Similar vertex
corrections appear in the calculation of other physical quantities 
such as the dynamic spin susceptibility.

\section{Discussion and Conclusions}
\label{sec:IV}
In this paper we have made a comparison of the spectral functions
of the $U=\infty$ Anderson model as calculated within the simplest
self--consistent conserving approximation, the NCA, and within the
NRG. At high energies $\omega>> T_{0}$ we found good quantitative agreement
for the auxiliary and physical spectral functions calculated within the
two methods in all parameter regimes. Some small discrepancies in the 
shape of high energy peaks could be attributed to the logarithmic 
discretization used in the NRG which
tends to give lower resolution and slight asymmetries to high energy
peaks. At lower energies we found good
quantitative agreement for both slave--boson and pseudo--fermion spectral
functions down to at least $T_{0}$. In the Kondo regime the agreement
between the NCA and NRG slave--boson spectral function $A_{b}^{+}$ extended
to well below the low energy scale $T_{0}=T_{K}$. Despite the accuracy
of the NCA auxiliary spectral functions down to $T_{K}$ we noted that the
impurity spectral density deviated from the essentially exact NRG result
on energy scales up to $5T_{K}$. The source of this discrepancy was traced
directly to the lack of the vertex part in the NCA expression for the impurity
spectral density, thus showing that vertex corrections are required for
the physical Green's functions even when the auxiliary particle dynamics
appears to be correctly described within the NCA (for energies down 
to $T_{0}$). In the Fermi liquid regime $\omega << T_{0}$, the NCA
gives results for the impurity spectral density in disagreement 
with exact results from Fermi liquid theory. It also gives the 
incorrect threshold exponents $\alpha_{f,b}$ for the auxiliary 
particle Green's functions. We emphasize, however, that although 
vertex corrections restore the Fermi liquid behaviour in the physical 
spectral functions and the correct low energy asymptotic behaviour of 
the auxiliary particle Green's functions, our results show that they 
are also needed for a correct description of the physical Green's 
functions at much higher energies $\omega\sim T_{0}$. This is the main 
conclusion of the comparisons we made.

We are in the process of including vertex corrections within a self--consistent
theory \cite{kroha.tmatrix}, thus going beyond the lowest order conserving 
approximation. The availability of accurate results for
dynamic properties via the NRG for impurity models makes the latter
a natural testing ground for developing such approximation schemes. 
These schemes could then be extended to study lattice models of
strongly correlated electrons, for which, at present, numerical 
renormalization group methods are not as well developed as for 
impurity models.

\acknowledgments
We are grateful to K.A. Muttalib and P. Hirschfeld for useful discussions
concerning self--consistent conserving approximations.
This work was supported by E.U. grant no. ERBCHRX CT93 0115 (TAC), 
the Alexander von Humboldt Foundation (JK) and the Deutsche 
Forschungsgemeinschaft (PW,TAC). 

\appendix
\section{Matrix Elements for Auxiliary Particle Operators}

In this appendix we give the expressions for the
matrix elements of the auxiliary particle operators, required for
the calculation of the spectral functions. The states, 
$|Q,N_{e},S,S_{z},r>_{N}$, of the 
Hamiltonian $H_{N}$ (which includes the orbitals 
$f_{\sigma},c_{0\sigma},\ldots,c_{N\sigma}$) are labeled
by the quantum numbers $Q,N_{e},S,S_{z}$ and an index $r$, where
$Q=\sum_{\sigma}f_{\sigma}^{\dagger}f_{\sigma}+b^{\dagger}b$ 
is the number of auxiliary
particles, $N_{e}$ is the total number of electrons, $S$ the total spin and
$S_{z}$ the z--component of the total spin and the index
$r$ distinguishes states with the same conserved quantum numbers. A product 
basis set $|Q,N_{e},S,S_{z},ri>_{N}$ for the subspace $(Q,N_{e},S,S_{z})$ 
of $H_{N}$ in terms of eigenstates of $H_{N-1}$ and states 
$|0>,|\uparrow>,|\downarrow>,|\uparrow\downarrow>$ of the orbital 
$c_{N\sigma}$ is defined by
\begin{eqnarray}
|Q,N_{e},S,S_{z},r,1>_{N} & = & |Q,N_{e},S,S_{z},r>_{N-1}|0>,\label{prod1}\\
|Q,N_{e},S,S_{z},r,2>_{N} & = & \sqrt{\frac{S+S_{z}}{2S}}
|Q,N_{e}-1,S-\frac{1}{2},S_{z}-\frac{1}{2},r>_{N-1}|\uparrow>\nonumber\\
                          & + & \sqrt{\frac{S-S_{z}}{2S}}
|Q,N_{e}-1,S-\frac{1}{2},S_{z}+\frac{1}{2},r>_{N-1}|\downarrow>,\label{prod2}\\
|Q,N_{e},S,S_{z},r,3>_{N} & = & - \sqrt{\frac{S-S_{z}+1}{2S+2}}
|Q,N_{e}-1,S+\frac{1}{2},S_{z}-\frac{1}{2},r>_{N-1}|\uparrow>\nonumber\\
                          & + & \sqrt{\frac{S+S_{z}+1}{2S+2}}
|Q,N_{e}-1,S+\frac{1}{2},S_{z}+\frac{1}{2},r>_{N-1}|\downarrow>,\label{prod3}\\
|Q,N_{e},S,S_{z},r,4>_{N} & = & |Q,N_{e}-2,S,S_{z},r>_{N-1}|\uparrow\downarrow>
\label{prod4}
\end{eqnarray}
The reduced matrix elements 
${_N}<Q,N_{e},S,r||f^{\dagger}||Q-1,N_{e}-1,S\pm 1/2,s>_{N}$ 
and ${_N}<Q,N_{e},S,r||b^{\dagger}||Q-1,N_{e},S,s>_{N}$ are 
calculated recursively following the recursive calculation of 
the matrix elements 
${_N}<Q,N_{e},S,r||c_{N}^{\dagger}||Q,N_{e}-1,S\pm 1/2,s>_{N}$ 
required for setting up the Hamiltonian $H_{N}$. 
Details of the latter
can be found in \cite{nrg.ref}. We follow the notation of \cite{nrg.ref} and
denote by $U_{Q,N_{e},S}^{N}(ri,p), p=1,\ldots,R_{Q,N_{e}S}^{N}$ the matrix
of eigenvectors of the subspace $(Q,N_{e}S,S_{z})$ of $H_{N}$ 
where $R_{Q,N_{e}S}^{N}$ is the 
dimensionality of this subspace and $r$ and $i=1,\ldots,4$ label the product 
state basis which is related to the diagonal basis by the unitary 
transformation 
\begin{equation}
|Q,N_{e},S,S_{z},p>_{N} = \sum_{r,i}U_{Q,N_{e},S}(p,ri)|r>_{N-1}|i>
\label{unitary-trans}
\end{equation}
with $|r>_{N-1}|i>$ denoting one of the four product states 
(\ref{prod1}--\ref{prod4}) defined above.
Defining,
\begin{eqnarray}
M_{Q,N_{e},S}^{f\pm,N}(r,r') & \equiv & _{N}<Q,N_{e},S,r||f^{\dagger}||
Q-1,N_{e}-1,S\pm\frac{1}{2},r'>_{N}\\
M_{Q,N_{e},S}^{b,N}(r,r') & \equiv & _{N}<Q,N_{e},S,r||
b^{\dagger}||Q-1,N_{e},S,r'>_{N},
\end{eqnarray}
and using the unitary transformation (\ref{unitary-trans}) we have,
\begin{eqnarray}
M_{Q,N_{e},S}^{f\pm,N}(p,q) &\equiv& 
{_{N}}<Q,N_{e},S,p||f^{\dagger}||
Q-1,N_{e}-1,S\pm\frac{1}{2}, q>_{N}\nonumber\\
& = & 
\sum_{ri,r'i'}
U_{Q,N_{e}S}^{N}(p,ri)U_{Q-1,N_{e}-1,S\pm\frac{1}{2}}^{N}(q,r'i')\nonumber\\
&\times&
 _{N-1}<Q,N_{e},S,ri||f^{\dagger}||Q-1,N_{e}-1,S\pm\frac{1}{2},r'i'>_{N-1}\\
M_{Q,N_{e},S}^{b,N}(p,q) & \equiv & 
_{N}<Q,N_{e},S,p||b^{\dagger}||Q-1,N_{e},S,q>_{N}\nonumber\\
 & = & \sum_{ri,r'i'}U_{Q,N_{e}S}^{N}(p,ri)U_{Q-1,N_{e},S}^{N}(q,r'i')
\nonumber\\
&\times& _{N-1}<Q,N_{e},S,ri||b^{\dagger}||Q-1,N_{e},S,r'i'>_{N-1}.
\end{eqnarray}
Evaluating these using the expressions (\ref{prod1}--\ref{prod4}) gives
an expression relating the matrix elements at iteration $N$ to those at
iteration $N-1$,

\begin{eqnarray*}
M_{Q,N_{e},S}^{f\pm,N}(p,q) 
& = & C^{\pm}_{11}\sum_{rr'} U_{Q,N_{e}S}^{N}(p,r1) U_{Q-1,N_{e}-1,S\pm\frac{1}{2}}^{N}(q,r'1)M_{Q,N_{e},S}^{f\pm,N-1}(r,r')\\
& + & C^{\pm}_{22}\sum_{rr'} U_{Q,N_{e}S}^{N}(p,r2) U_{Q-1,N_{e}-1,S\pm\frac{1}{2}}^{N}(q,r'2)M_{Q,N_{e}-1,S-\frac{1}{2}}^{f\pm,N-1}(r,r')\\
& + & C^{\pm}_{33}\sum_{rr'} U_{Q,N_{e}S}^{N}(p,r3) U_{Q-1,N_{e}-1,S\pm\frac{1}{2}}^{N}(q,r'3)M_{Q,N_{e}-1,S+\frac{1}{2}}^{f\pm,N-1}(r,r')\\
& + & C^{\pm}_{44}\sum_{rr'} U_{Q,N_{e}S}^{N}(p,r4) U_{Q-1,N_{e}-1,S\pm\frac{1}{2}}^{N}(q,r'4)M_{Q,N_{e}-2,S}^{f\pm,N-1}(r,r')\\
& + & C^{\pm}_{23}\sum_{rr'} U_{Q,N_{e}S}^{N}(p,r2) U_{Q-1,N_{e}-1,S\pm\frac{1}{2}}^{N}(q,r'3)M_{Q,N_{e}-1,S\pm\frac{1}{2}}^{f\mp,N-1}(r,r')\\
& + & C^{\pm}_{32}\sum_{rr'} U_{Q,N_{e}S}^{N}(p,r3) U_{Q-1,N_{e}-1,S\pm\frac{1}{2}}^{N}(q,r'2)M_{Q,N_{e}-1,S\pm\frac{1}{2}}^{f\mp,N-1}(r,r'),
\end{eqnarray*}
and,
\begin{eqnarray*}
M_{Q,N_{e},S}^{b,N}(p,q) 
& = & \sum_{rr'} U_{Q,N_{e}S}^{N}(p,r1) U_{Q-1,N_{e},S}^{N}(q,r'1)
M_{Q,N_{e},S}^{b,N-1}(r,r')\\
& + & \sum_{rr'} U_{Q,N_{e}S}^{N}(p,r2) U_{Q-1,N_{e},S}^{N}(q,r'2)
M_{Q,N_{e},S-\frac{1}{2}}^{b,N-1}(r,r')\\
& + & \sum_{rr'} U_{Q,N_{e}S}^{N}(p,r3) U_{Q-1,N_{e},S}^{N}(q,r'3)
M_{Q,N_{e}-1,S+\frac{1}{2}}^{b,N-1}(r,r')\\
& + & \sum_{rr'} U_{Q,N_{e}S}^{N}(p,r4) U_{Q-1,N_{e},S}^{N}(q,r'4)
M_{Q,N_{e}-2,S}^{b,N-1}(r,r'),
\end{eqnarray*}
where the coefficients $C_{ii'}^{\pm}$ are given in Table \ref{table2}.

\section{Errors and Sum Rules}

In this appendix we outline some of the checks carried out to ensure 
the correctness of the numerical renormalization group programs. 
The eigenvectors and eigenvalues of $H_{0}$ can be calculated analytically
and these can then be used to set up the matrices for $N=1$. The latter
have been compared with those generated by the computer programs for $N=1$
and found to be identical. We have also checked the recursive 
evaluation of the matrix elements, $<r|f_{\sigma}^{\dagger}|s>, 
<r|b^{\dagger}|s>$ and $<r|c_{N\sigma}^{\dagger}|s>$ by making use of the
commutation relations for the creation and annihilation operators appearing in
the Hamiltonian (\ref{eq:AM}),
\begin{eqnarray}
f_{\sigma}f_{\sigma}^{\dagger} + f_{\sigma}^{\dagger}f_{\sigma} & = & 1\label{com1},\\
b_{\sigma}b_{\sigma}^{\dagger} - b_{\sigma}^{\dagger}b_{\sigma} & = & 1\label{com2},\\
c_{N\sigma}c_{N\sigma}^{\dagger} + c_{N\sigma}^{\dagger}c_{N\sigma} & = & 1\label{com3}.
\end{eqnarray}
For any state $|Q,N_{e},S,S_{z},k>_{N}$ in the Hilbert space of $H_{N}$, the
completeness relation $\sum_{r}|r><r|=1$ together with 
(\ref{com1}--\ref{com3}) yields,

\begin{eqnarray}
1 & = & \sum_{k'}
|<Q,N_{e},S,k||f^{\dagger}||Q-1,N_{e}-1,S-\frac{1}{2},k'>|^{2}\nonumber\\
& + & \frac{1}{2S+2}\sum_{k'}
|<Q,N_{e},S,k||f^{\dagger}||Q-1,N_{e}-1,S+\frac{1}{2},k'>|^{2}\nonumber\\
& + & \sum_{k'}
|<Q+1,N_{e}+1,S+\frac{1}{2},k'||f^{\dagger}||Q,N_{e},S,k>|^{2},\label{eq:sum-rule1}\\
1 & = & \sum_{k'}
|<Q+1,N_{e},S,k'||b^{\dagger}||Q,N_{e},S,k>|^{2}\nonumber\\
& - & \sum_{k'}
|<Q,N_{e},S,k||b^{\dagger}||Q-1,N_{e},S,k'>|^{2},\label{eq:sum-rule2}\\
1 & = & \sum_{k'}
|<Q,N_{e},S,k||c_{N}^{\dagger}||Q,N_{e}-1,S-\frac{1}{2},k'>|^{2}\nonumber\\
& + & \frac{1}{2S+2}\sum_{k'}
|<Q,N_{e},S,k||c_{N}^{\dagger}||Q,N_{e}-1,S+\frac{1}{2},k'>|^{2}\nonumber\\
& + & \sum_{k'}
|<Q,N_{e}+1,S+\frac{1}{2},k'||c_{N}^{\dagger}||Q,N_{e},S,k>|^{2}\label{eq:sum-rule3}.
\end{eqnarray}
In the calculations we verified that these relations were satisfied 
to within rounding errors for each state $|Q,N_{e},S,k>_{N}$ in 
$H_{N}$ for $N=0,1,\ldots,4$, when all states are retained. This
gives a reliable test of the formulae in Appendix~A and on the routines for the
recursive evaluation of the matrix elements. We note that once higher 
energy states start being eliminated, typically for $N > 4$, the LHS 
of the above expressions  will be less than unity due to the missing 
states. The sum rule (\ref{eq:sum-rule3})
also provides a check on the construction of the Hamiltonian $H_{N}$, since
the latter depends on the matrix elements 
$<Q=1,N_{e}+1,S\pm\frac{1}{2},k'||c_{N}^{\dagger}||Q=1,N_{e},S,k>$ 
\cite{nrg.ref}. The above tests on all quantities appearing in the iterative
procedure for the first few iterations virtually eliminates the possibility
of errors.

\section{Exact Projection onto the Physical Subspace}
In order to effect the constraint of the dynamics to the physical 
Hilbert subspace it is convenient to add the term $\lambda Q$ to the 
Hamiltonian, where $-\lambda $ is a ``chemical potential'' 
associated with the auxiliary particle 
number $Q$. The operator constraint $Q=1$ is imposed exactly on the
expectation value of any operator $O$ by differentiating with 
respect to the fugacity $\zeta = e^{-\beta\lambda}$ and then taking the
limit $\lambda \rightarrow \infty$ \cite{coleman.84}: 
\begin{equation}
\langle O \rangle _C = \lim _{\lambda \rightarrow \infty}
\frac {\frac{\partial }{\partial \zeta} \mbox{tr}
       \bigl[ O e^{-\beta (H+\lambda Q)} \bigr]} 
      {\frac{\partial }{\partial \zeta} \mbox{tr}
       \bigl[ e^{-\beta (H+\lambda Q)} \bigr]},
\label{eq:projection}
\end{equation}
where the trace is taken over the complete, enlarged Hilbert space.
In particular, we state the following two results which
are of use to us in this paper and which follow straightforwardly
from the above (for details see \cite{coleman.84}). First,
the canonical partition function in the subspace $Q=1$ is
\begin{eqnarray}
Z_C & = & \lim _{\lambda \rightarrow \infty} \mbox{tr}
          \bigl[ Q e^{-\beta (H+\lambda (Q-1))} \bigr] \nonumber \\ 
    & = & \lim _{\lambda \rightarrow \infty}
          \bigl( e^{\beta\lambda} \langle Q \rangle _{GC}(\lambda ) 
          \bigr)Z_{Q=0},\label{eq:zcan}
\end{eqnarray}
where the subscripts $GC$ and $C$ denote the grand--canonical and the
canonical ($Q=1$) expectation value, respectively. Second, the 
canonical $Q=1$ expectation
value of any operator $O$ having a  zero expectation value in the $Q=0$
subspace is given by,
\begin{equation}
\langle O \rangle _C = \lim _{\lambda \rightarrow \infty}
\frac {\langle O \rangle _{GC}(\lambda )}{ 
\langle Q \rangle _{GC}(\lambda )}
\label{eq:canonical-expectation}
\end{equation}

Thus, we obtain the
constrained $d$--electron Green function in terms of the grand--canonical
one ($G_{d}(\omega ,T,\lambda )$) as
\begin{equation} 
G_d(\omega ) = \lim _{\lambda \rightarrow \infty}
\frac {G_{d}(\omega ,T,\lambda )}{\langle Q \rangle _{GC} (\lambda )} 
\label {eq:Gd}
\end{equation}
In the enlarged Hilbert space ($Q=0,1,2,...$) 
$G_{d}(\omega, T, \lambda )$ may be 
expressed in terms of the pseudo--fermion and slave boson Green functions
using Wick's theorem. It then follows from Eq.~(\ref{eq:Gd}) that the
operator constraint $Q=1$ is imposed on the auxiliary Green's functions by
simply taking the limit $\lambda \rightarrow \infty$ of the respective
unconstrained functions. Clearly, by this procedure all excitation
energies of pseudo--fermions and slave bosons are shifted to $\infty $.
It is therefore convenient to re--define the auxiliary particle frequency
scale as $\omega \rightarrow \omega + \lambda $ before taking the limit
$\lambda \rightarrow \infty$. Note that this does not affect the energy 
scale of physical quantities (like the local $d$ electron Green's function), 
which is the {\it difference} between the 
the pseudo--fermion and the slave--boson energy.

\section{NCA Calculation of the Auxiliary Spectral Functions $A_{f,b}^{\pm}$}

In order to enter the asymptotic power law regime of the auxiliary spectral
functions and to compare with the T=0 results of the NRG, the NCA must be
evaluated for temperatures several orders of magnitude below $T_0$, the
low temperature scale of the model. The equations are solved numerically
by iteration. In this appendix the two main procedures are described to 
make the diagrammatic auxiliary particle technique suitable for the lowest 
temperatures.

The grand--canonical expectation value of the auxiliary particle number
appearing in Eq.~(\ref{eq:Gd}) is given in terms of the (unprojected) 
auxiliary particle spectral functions $A_{f,b}^{+}(\omega,T,\lambda)$ by,

\begin{equation}
\langle Q \rangle _{GC} (\lambda ) = \int \mbox{d}\omega
\bigl[ f(\omega) \sum _{\sigma } A_{f\sigma }^+ (\omega,T,\lambda)
+ b(\omega) A_b^+(\omega,T,\lambda)
\bigr]
\end{equation}
where $f(\omega)$, $b(\omega)$ denote the Fermi and Bose functions. 
Substituting this into the expression (\ref{eq:zcan}) for the 
canonical partition function we obtain after carrying out
the transformation $\omega \rightarrow \omega +\lambda $, and taking
the limit $\lambda \rightarrow \infty $
\begin{eqnarray}
e^{-\beta F_{imp}(T)} & = & 
\frac{Z_{C}}{Z_{Q=0}}  = 
\lim_{\lambda\rightarrow\infty}e^{\beta\lambda}
\langle Q \rangle_{GC}(\lambda) \nonumber\\
& = & \int \mbox{d}\omega e^{-\beta\omega} \bigl[  \sum _{\sigma } 
A_{f\sigma }^+(\omega,T) + A_b^+(\omega,T)\bigr],
\label{eq:Qproj}
\end{eqnarray}
where $A_{f,b}^{+}(\omega,T)$ are now the projected spectral functions
as defined in Eq.~(\ref{eq:aplus}), and by definition 
$F_{imp}=-\frac{1}{\beta}\ln(Z_{C}/Z_{Q=0})$ is the impurity 
contribution to the Free energy.

The numerical evaluation of expectation values like 
$\langle Q\rangle _{GC}(\lambda\rightarrow\infty)$ (Eq.~(\ref{eq:Qproj}))
or $\Sigma_{c\sigma}(\omega,T,\lambda\rightarrow\infty)$ 
(Eq.~(\ref{eq:nca-selfc})) is non--trivial
(1) because at $T=0$ the auxiliary spectral functions $A_{f,b}^+(\omega,T)$
are divergent at the threshold frequency $E_{0}$, where the exact
position of $E_{0}$ is a priori not known, and (2) because 
the Boltzmann factors $e^{-\beta\omega}$ diverge strongly for $\omega < 0$.
Therefore, we apply the following transformations:

(1) Before performing the projection $\omega \rightarrow \omega +\lambda$,
$\lambda\rightarrow\infty$ we re--define the frequency scale of all
auxiliary particle functions $A_{f,b}^{\pm}$ according to  
$\omega \rightarrow \omega +\lambda_0$, where $\lambda_0$ is a finite
parameter.  
In each iteration $\lambda _0$  is then determined  such that
\begin{equation}
\int \mbox{d}\omega
e^{-\beta\omega} \bigl[ \sum _{\sigma } A_{f\sigma }^+(\omega) +
A_b^+(\omega)\bigr]=1
\label{eq:fixl}
\end{equation}
where $A_{f,b}^{+}(\omega)=\lim_{\lambda\rightarrow\infty}
A_{f,b}^{+}(\omega+\lambda_0+\lambda,T,\lambda)$ 
is now an auxiliary spectral function with the new reference energy.
It is seen from Eq.~(\ref{eq:Qproj}) that 
$\lambda_0(T)=F_{imp}(T) = F_{Q=1}(T)-F_{Q=0}(T)$, i.e. $\lambda_0$ is the
chemical potential for the auxiliary particle number $Q$, or equivalently
the impurity contribution to the Free energy. The difference of the 
Free energies becomes equal to the threshold energy 
$E_{0}=E_{Q=1}^{GS}-E_{Q=0}^{GS}$ at
$T=0$, so the energy scale of the shifted spectral functions defined 
above coincides exactly with that of the NRG spectral functions defined in
(\ref{eq:aplus}). More importantly, however, the above way of determining a 
``threshold'' is less ad hoc than, for example, defining it by a maximum
in some function appearing in the NCA equations. 
It is also seen from Eq.~(\ref{eq:fixl}) that this procedure 
defines the frequency scale
of the auxiliary particles such that the $T=0$ threshold divergence of the
spectral functions is at the {\it fixed} frequency $\omega = 0$.
This substantially increases the precision as well as the speed of numerical
evaluations. Eq. (\ref{eq:Gd}) for the projected $d$ electron Green's 
function becomes
\begin{equation}
G_d(\omega ) = \lim _{\lambda \rightarrow \infty} e^{\beta\lambda}
G_{d}(\omega,T,\lambda ).
\end{equation}
  
(2) The divergence of the Boltzmann factors implies that the 
self--consistent solutions for $A_{f,b}^+(\omega)$
vanish exponentially $\sim e^{\beta\omega}$ for negative frequencies.
It is convenient, not to formulate the self--consistent equations in terms of
$A_{f,b}^{\pm}$ like in earlier evaluations \cite{bickers.87}, but to define
new functions $\tilde A_{f,b}(\omega)$ and 
$\mbox{Im}\tilde\Sigma_{f,b}(\omega)$ such that
\begin{eqnarray}
A_{f,b}^+(\omega) & = & f(-\omega )~ \tilde A_{f,b}(\omega)\\
\mbox{Im}\Sigma_{f,b}(\omega) & = & f(-\omega ) ~ \mbox{Im} 
\tilde\Sigma_{f,b}(\omega).
\end{eqnarray}
After fixing the chemical potential $\lambda _0$ and performing the
projection onto the physical subspace, the canonical partition function
(Eq.~(\ref{eq:zcan})) behaves as 
$\lim _{\lambda\rightarrow\infty}e^{\beta(\lambda-\lambda_0)}\;Z_C(T) =1$, 
and it
follows immediately from the definition of $A_{f,b}^{-}$ that
\begin{equation}
A_{f,b}^-(\omega)=f(\omega ) ~ \tilde A_{f,b}(\omega).
\end{equation}
In this way all exponential divergencies are absorbed by one single function
for each particle species.
The NCA equations in terms of these functions are free of divergencies of the
statistical factors and read
\begin{eqnarray}
\mbox{Im}\tilde\Sigma_{f\sigma}(\omega -i0,T) 
& = & V^{2}\sum_{k}\frac{(1-f(\epsilon_{k}))(1-f(\omega -\epsilon _{k}))}
{1-f(\omega )}
\tilde A_{b}(\omega-\epsilon_{k})\\
\mbox{Im}\tilde\Sigma_{b}(\omega -i0,T) 
& = & V^{2}\sum_{k,\sigma}\frac{f(\epsilon_{k})(1-f(\omega +\epsilon _{k}))}
{1-f(\omega )}
\tilde A_{f\sigma}(\omega+\epsilon_{k}),\\
\langle Q \rangle  (\lambda _0,\lambda\rightarrow\infty) & = &
\int \mbox{d}\omega
f(\omega)\bigl[ \sum _{\sigma } \tilde A_{f\sigma }(\omega) +
\tilde A_b (\omega)\bigr] = 1\\
\mbox{Im}G_{d\sigma}(\omega -i0,T) & = & 
\int \mbox{d}\epsilon [f(\epsilon +\omega)f(-\epsilon )+
f(-\epsilon -\omega)f(\epsilon)]
\tilde A_{f\sigma}(\epsilon +\omega)\tilde A_b(\epsilon),
\end{eqnarray}
where the real parts of the 
self--energies $\Sigma _f$, $\Sigma _b$, $\Sigma _c$ 
are determined from a Kramers--Kronig relation, and the auxiliary spectral 
functions
are the imaginary parts of the Green's functions,
$A_{f\sigma}^+(\omega)=-\mbox{Im}\biggl[\bigl(\omega+\lambda_0-i0-\epsilon_d-
\Sigma_{f\sigma}(\omega-i0)\bigr)^{-1}\biggr]$, 
$A_b^+(\omega)=-\mbox{Im}\biggl[\bigl( \omega+\lambda_0-i0-\Sigma_b(\omega-i0) 
\bigr)^{-1}\biggr]$.

The above method allows to solve the NCA equations effectively
for temperatures down to typically $T=10^{-4}T_0$. It may be shown that
the same procedure can also be applied to self--consistently compute vertex
corrections \cite{kroha.privatecommun.95} beyond the NCA.


\newpage
\begin{figure}
\caption{(a) The lowest order contribution to the generating functional $\Phi$,
and, (b), the renormalized vertex part 
$\Lambda(i\omega_{\nu},i\omega_{\nu}-i\omega)$,
entering the expression ({\protect{\ref{eq:realgf+vertex}}})
for the d--electron Green's function (with frequency conservation we can 
omit $i\omega_{n}$ in $\Lambda$).
The solid lines are for band electrons, the dashed lines are for 
pseudo--fermions and the wiggly lines are for slave--bosons.
}
\label{vertex-part}
\end{figure}
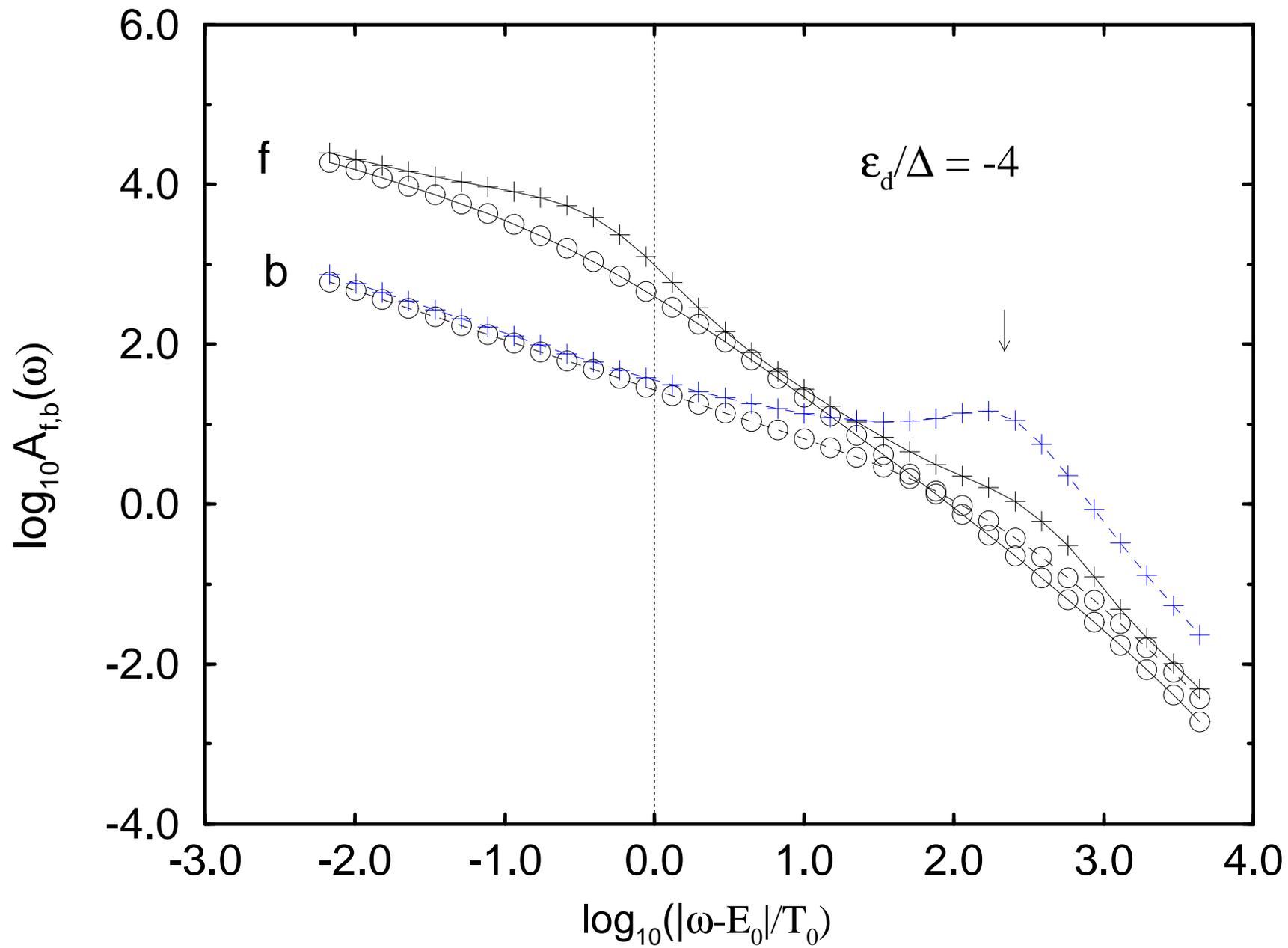
\begin{figure}
\caption{
The $T=0$ NRG pseudo--fermion $A_{f}^{\pm}$ (solid lines) and slave--boson 
$A_{b}^{\pm}$ (dashed lines)
spectral functions in the Kondo case $\epsilon_{d}/\Delta=-4$, 
$T_{0}/\Delta = 1.87\times 10^{-2}$, $n_{d}=0.874$. The $+$ signs
are for the spectral function above the threshold, $E_{0}$, 
and the circles are for the spectral function below the threshold.
The arrow indicates the position of $|\epsilon_{d}|$.}
\label{sd-kondo}
\end{figure}
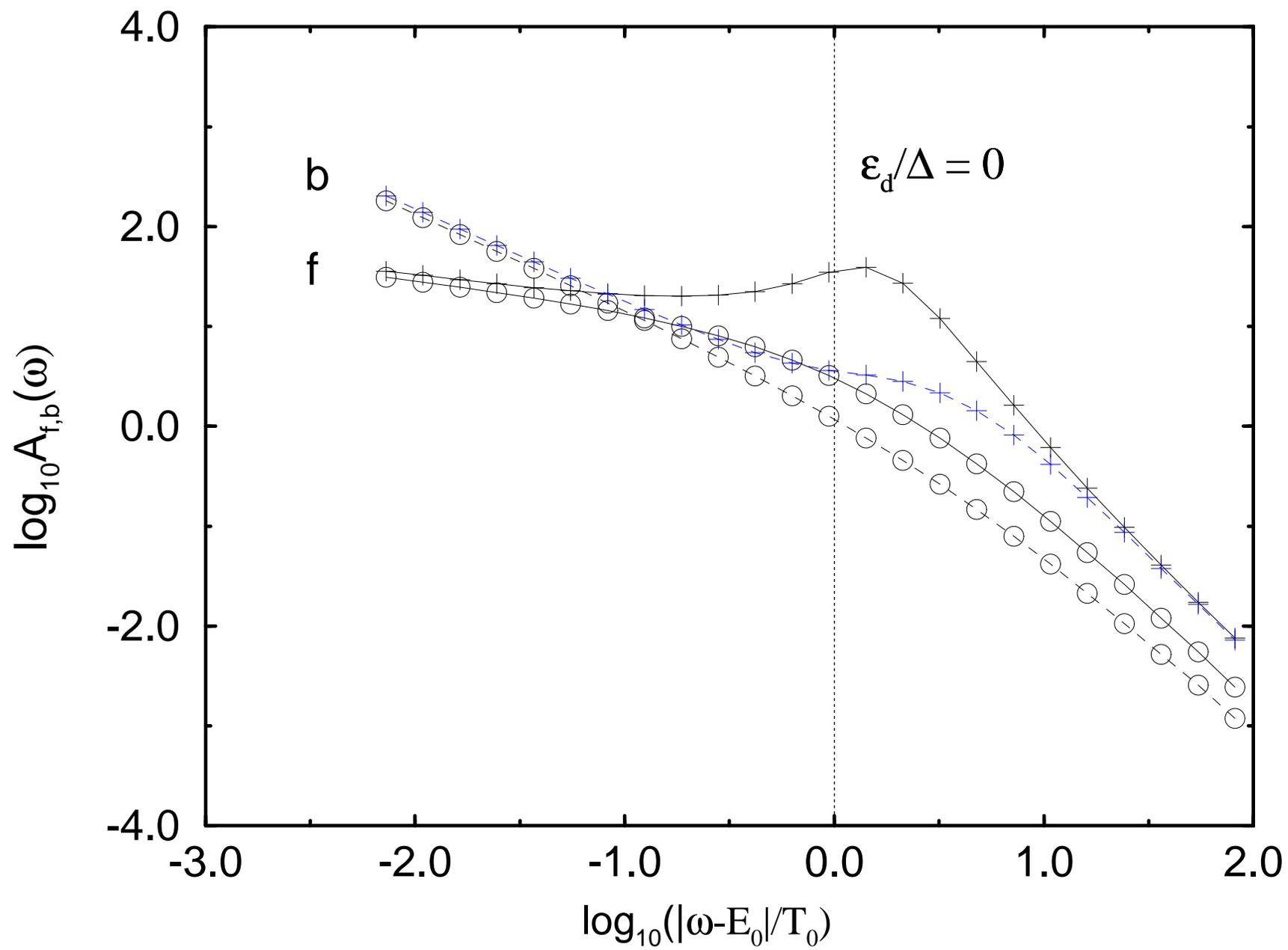
\begin{figure}
\caption{
The $T=0$ NRG spectral functions $A_{f,b}^{\pm}$ 
in the mixed valent regime $\epsilon_{d}/\Delta=0$, $n_{d}=0.314$ 
with notation as in Fig.\ {\protect\ref{sd-kondo}}. }
\label{sd-mixed-valent}
\end{figure}
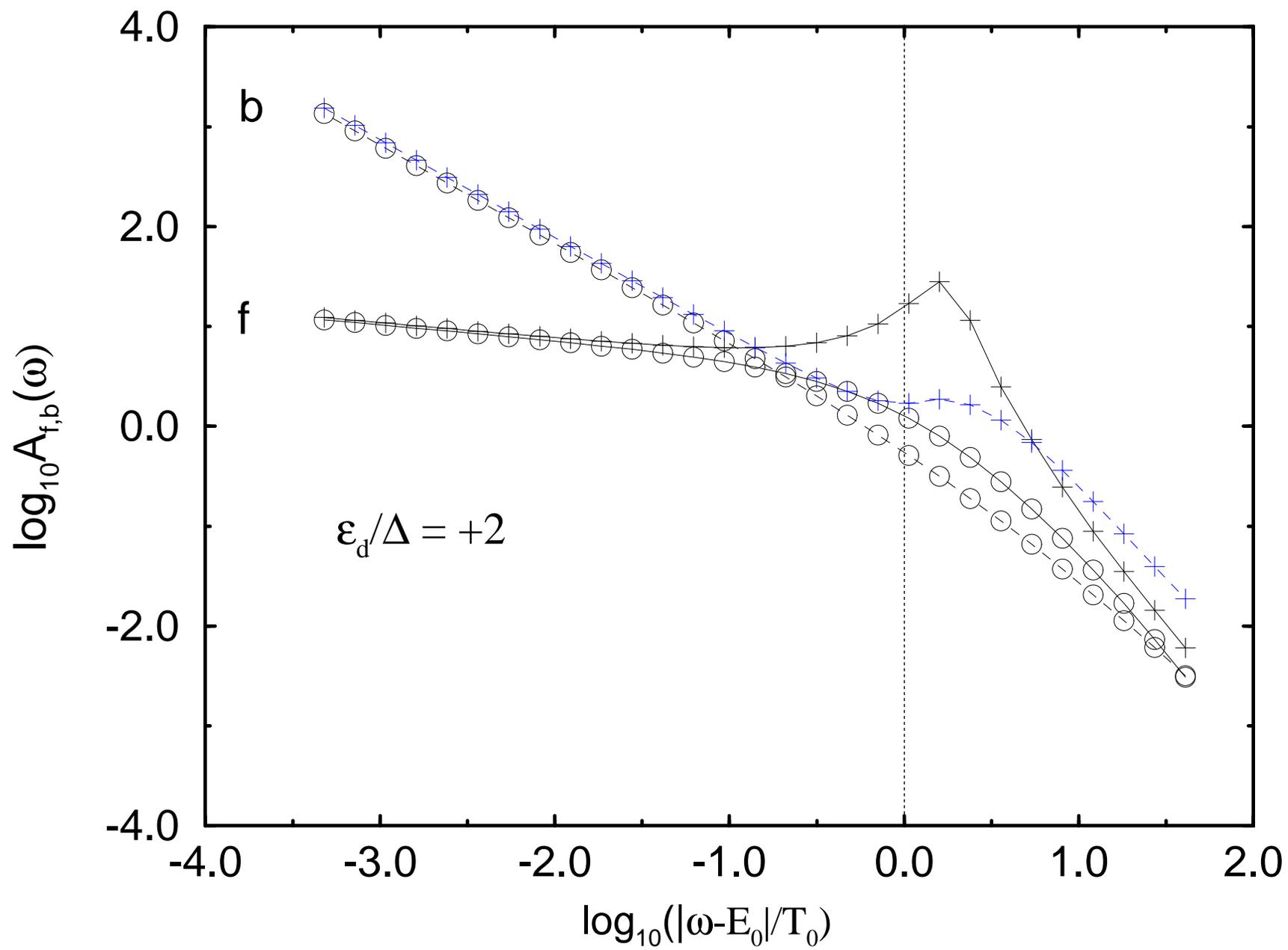
\begin{figure}
\caption{
The $T=0$ NRG spectral functions $A_{f,b}^{\pm}$ 
in the empty orbital regime $\epsilon_{d}/\Delta=+2$, $n_{d}=0.172$, 
with notation as in Fig.\ {\protect\ref{sd-kondo}}. }
\label{sd-empty}
\end{figure}
\begin{figure}
\caption{
The exponents $\alpha_{f}$ ($\diamond$), $\alpha_{b}$ ($\circ$) 
deduced from the asymptotic power
law behaviour of the auxiliary spectral functions as calculated within
the NRG for different values of the occupation $n_{d}$. The solid
lines are the functions $n_{d}-n_{d}^{2}/2$ and $1-n_{d}^{2}/2$.
}
\label{exponents}
\end{figure}
\begin{figure}
\caption{
The temperature dependence of the occupation number $n_{d}(T)$ for
different $\epsilon_{d}$ (the curves are labeled by $\epsilon_{d}/\Delta$).
The high temperature limit of $2/3$ (indicated by an arrow) is
recovered in all cases.
}
\label{valence}
\end{figure}
\begin{figure}
\caption{
Comparison of the NRG ($\diamond $) and NCA ($\circ $) auxiliary spectral
functions $A_{f,b}^{+}$ (a--b), $A_{f,b}^{-}$ (c--d) in the Kondo case
$\epsilon_{d}/\Delta =-4$. The arrow indicates the position 
of $|\epsilon_{d}|$. The NRG results are for $T=0$ and the NCA results
are for $T=1.0\times 10^{-6}D$. The divergence of the NCA spectral functions
for $\omega\rightarrow 0$ is cut off below the finite 
temperature $T<<T_{0}$.
}
\label{sd-kondo-nca}
\end{figure}
\begin{figure}
\caption{
Comparison of the NRG ($\diamond$) and NCA ($\circ$) auxiliary spectral
functions $A_{f,b}^{+}$ (a--b), $A_{f,b}^{-}$ (c--d) in the mixed valent case
$\epsilon_{d}/\Delta =0$. The NRG results are for $T=0$ and the NCA results
are for $T=1.0\times 10^{-6}D$.
}
\label{sd-mixed-valent-nca}
\end{figure}
\begin{figure}
\caption{
Comparison of the NRG ($\diamond$) and NCA ($\circ$) auxiliary spectral
functions $A_{f,b}^{+}$ (a--b), $A_{f,b}^{-}$ (c--d) in the empty orbital 
case $\epsilon_{d}/\Delta =+2$. The NRG results are for $T=0$ and the 
NCA results are for $T=1.0\times 10^{-6}D$.
}
\label{sd-empty-nca}
\end{figure}
\begin{figure}
\caption{
The impurity spectral function in the Kondo regime
$\epsilon_{d}=-4\Delta$, $T_{0}/\Delta=1.87\times 10^{-2}$,
for (a) high energies and, the low energy region (b) $-5 \le \omega/T_{0} \le 10$, and (c) $-2 \le \omega/T_{0} \le +2$. The dashed curve is the NRG result 
$\rho_{d}^{NRG}$, the dot--dashed curve is the NRG result without the
vertex part in ({\protect\ref{eq:nrg-convolution}}), and the solid
curve is the NCA result. The NRG results are for $T=0$ and the 
NCA results are for $T=1.0\times 10^{-6}D$.
}
\label{rsd-kondo}
\end{figure}
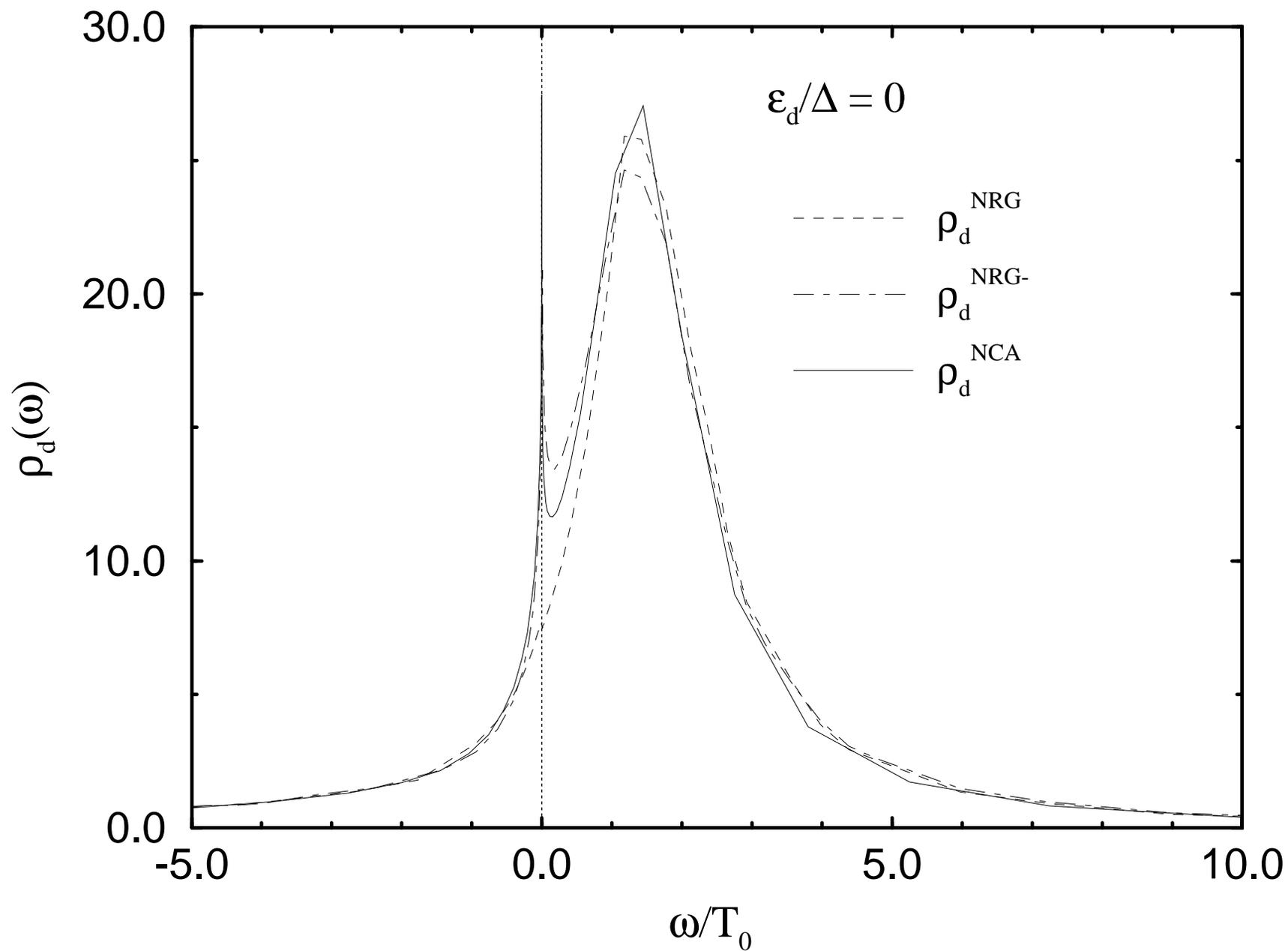
\begin{figure}
\caption{
The impurity spectral function in the mixed valent regime
$\epsilon_{d}/\Delta=0$, with notation 
as in Fig.\ {\protect\ref{rsd-kondo}}. The NRG results are for $T=0$ and the 
NCA results are for $T=1.0\times 10^{-6}D$.
}
\label{rsd-mixed-valent}
\end{figure}
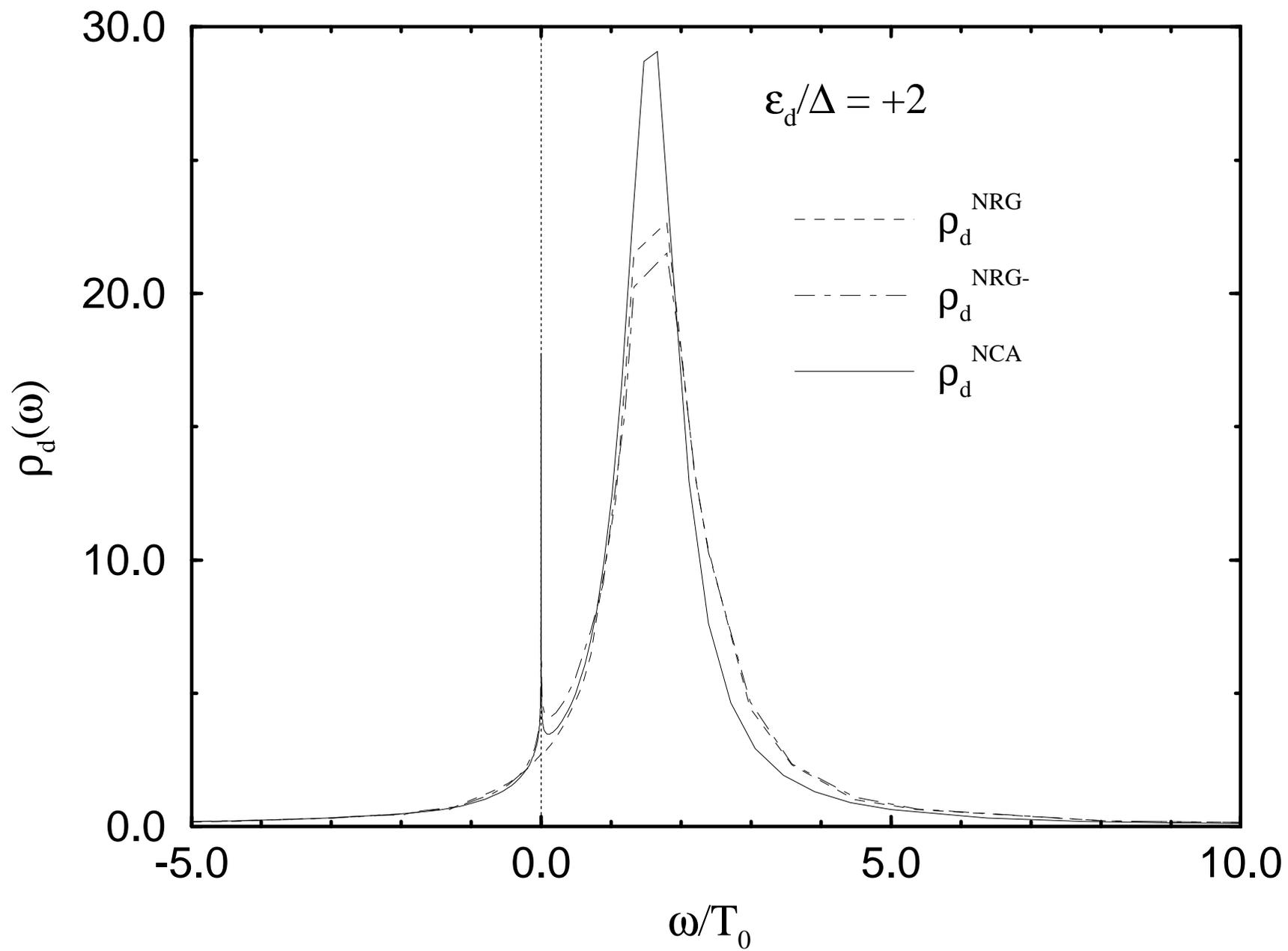
\begin{figure}
\caption{
The impurity spectral function in the empty orbital regime
$\epsilon_{d}/\Delta=+2$, with notation as in 
Fig.\ {\protect\ref{rsd-kondo}}. The NRG results are for $T=0$ and the 
NCA results are for $T=1.0\times 10^{-6}D$.
}
\label{rsd-empty}
\end{figure}
\newpage
\protect\begin{table}
\caption{
The threshold exponents $\alpha_{f,b}$ for the auxiliary spectral 
functions $A_{f,b}^{\pm}\sim |\omega-E_{0}|^{-\alpha_{f,b}}$. The
quantities $\alpha_{f,b}'$ are $n_{d}-n_{d}^{2}/2$ and $1-n_{d}^{2}/2$,
respectively where $n_{d}=n_{d}^{NRG}$ is the impurity occupation 
calculated from the NRG partition function (quantities
shown to 3 significant figures). The NCA results for the impurity
occupation, $n_{d}^{NCA}$, are also shown. The impurity spectral 
density at the Fermi level calculated within the NCA, $\rho_{d}^{NCA}(\epsilon_{F})$,  and NRG, $\rho_{d}^{NRG}(\epsilon_{F})$, are tabulated 
together with the $\%$ relative error
in the Friedel sum rule, 
$\rho_{d}(\epsilon_{F})=\sin^{2}(\pi n_{d}/2)/\pi\Delta$. 
The exponent $\alpha_{f}$ is
difficult to estimate close to $n_{d}=1$ due to the small
Kondo scale and the slow asymptotic behaviour of the pseudo--fermion 
spectral function. The low energy scale $T_{0}$ is $T_{K}$ given by
(\protect{\ref{eq:kondo-temp}}) in the Kondo regime, 
$\Delta$ in the mixed valent
regime and $\epsilon_{d}$ in the empty orbital regime.
}
\label{table1}
\begin{tabular}{cccccccccc}
$\epsilon_{d}/\Delta$
& $n_{d}^{NRG}$ 
& $n_{d}^{NCA}$ 
& $\rho_{d}^{NRG}(\epsilon_{F})$ 
& $\rho_{d}^{NCA}(\epsilon_{F})$ 
& $T_{0}/\Delta$ 
& $\alpha_{f}$
& $\alpha_{f}'$
& $\alpha_{b}$
& $\alpha_{b}'$\\
\tableline
$-7$ & 0.947  &$-$&$-$&$-$& $1.68\times 10^{-4}$ & 0.501 &  0.499 & 0.552 & 0.552\\
$-6$ & 0.934  &$-$&$30.7-2.5\%$&$-$& $8.07\times 10^{-4}$ & 0.499 &  0.498 & 0.563 & 0.564\\
$-5$ & 0.913  &$0.909$&$29.7-5.0\%$&$31.1-0.3\%$& $3.88\times 10^{-3}$ & 0.499 &  0.496 & 0.583 & 0.583\\
$-4$ & 0.874  &$0.865$&$30.1-1.7\%$&$34.3+12.1\%$& $1.87\times 10^{-2}$ & 0.493 &  0.492 & 0.618 & 0.619\\
$-3$ & 0.796  &$0.781$&$27.8-3.3\%$&$30.4+7.8\%$& $8.98\times 10^{-2}$ & 0.480 &  0.479 & 0.684 & 0.683\\
$-2$ & 0.648  &$0.641$&$23.4+1.4\%$&$32.4+42.3\%$& $4.32\times 10^{-1}$ & 0.439 &  0.438 & 0.790 & 0.790\\
$-1$ & 0.460  &$0.464$&$13.5-3\% $&$28.5+102\%$& $1$                  & 0.354 &  0.354 & 0.894 & 0.894\\
$0$  & 0.314  &$0.322$&$7.06-1\%$&$26.3+252\%$& $1$                  & 0.265 &  0.265 & 0.951 & 0.951\\
$+1$ & 0.226  &$0.232$&$3.80-1\%$&$25.7+535\%$& $1$                  & 0.200 &  0.200 & 0.975 & 0.974\\
$+2$ & 0.172  &$0.176$&$2.27-0.1\%$&$17.5+637\%$& $2$                  & 0.158 &  0.157 & 0.985 & 0.985\\
\end{tabular}
\end{table}
\begin{table}
\caption{Coefficients $C_{ii'}^{\pm}$}
\label{table2}
\begin{tabular}{ccc}
$ii'$ & $C_{ii'}^{+}$ & $C_{ii'}^{-}$ \\
\tableline
$11$ & $1$ & $1$ \\
$22$ & $\sqrt{\frac{2S(2S+2)}{(2S+1)(2S+1)}}$ & $1$ \\
$33$ & $1$ & $\sqrt{\frac{2S(2S+2)}{(2S+1)(2S+1)}}$ \\
$44$ & $1$ & $1$ \\
$23$ & $0$ & $\frac{1}{2S+1}$ \\
$32$ & $-\frac{1}{2S+1}$ & $0$ \\
\end{tabular}
\end{table}

\begin{references}
\bibitem{anderson.87} P. W. Anderson, Science {\bf 235}, 1196 (1987).
\bibitem{fascinating-ideas} N. Nagaosa and P. A. Lee, {\em Phys. Rev.
Lett.} {\bf 64}, 2450 (1990); C. M. Varma, P. B. Littlewood, S.
Schmitt--Rink, E. Abrahams, and A. E. Ruckenstein, {\em Phys. Rev.
Lett.} {\bf 63}, 1996 (1987); P. W. Anderson, {\em Phys. Rev. Lett.}
{\bf 65}, 2306 (1990).
\bibitem{barnes.76} S. E. Barnes, J. Phys. {\bf F6}, 1375 (1976); {\bf
F7}, 2637 (1977).
\bibitem{baskaran.88} G. Baskaran and P. W. Anderson, Phys. Rev. B{\bf 37},
580 (1988); L. Ioffe and A. Larkin, Phys. Rev. B{\bf 39} 8988 (1989); 
N. Nagaosa and P. A. Lee, {\em Phys. Rev.} B{\bf 45}, 966 (1992).
\bibitem{kadanoff.61} G.~Baym and L.P.~Kadanoff, Phys.~Rev. {\bf 124},
287 (1961); G.~Baym, Phys.~Rev. {\bf 127} 1391 (1962).
\bibitem{hewson.93} A. C. Hewson, {\em The Kondo Problem to Heavy
Fermions}, C.U.P (1993) gives an extensive discussion of the magnetic
impurity problem.
\bibitem{coleman.84} P. Coleman, {\em Phys. Rev.} {\bf B29}, 3035 (1984).
\bibitem{bickers.87} N. E. Bickers, {\em Rev. Mod. Phys.} {\bf 59},
845 (1987); N. E. Bickers, D. L. Cox \& J. W. Wilkins, 
{\em  Phys. Rev.}  {\bf  B36}, 2036 (1987).
\bibitem{keiter.nca} H. Keiter and J. C. Kimball, {\em J. Appl. Phys.}
{\bf 42}, 1460 (1971); N. Grewe and H. Keiter, {\em Phys. Rev. B} {\bf
24}, 4420 (1981); 
\bibitem{kuramoto.nca} Y. Kuramoto, {\em Z. Phys.} {\bf B53}, 37 (1983);
H. Kojima, Y. Kuramoto and M. Tachiki, {\em Z. Phys.} {\bf B54}, 293 (1984); 
Y. Kuramoto and H. Kojima, {\em Z. Phys.} {\bf B57}, 95 (1984); 
Y. Kuramoto, {\em Z. Phys.} {\bf B65}, 29 (1986).
\bibitem{mh.84} E. M\"uller--Hartmann, {\em Z. Phys.} {\bf B57}, 281 (1984).
\bibitem{costi.94} T. A. Costi, P. Schmitteckert, J. Kroha and 
P. W\"{o}lfle {\em Phys. Rev. Lett.}{\bf 73}, 1275 (1994).
\bibitem{costi.94b} T. A. Costi, P. Schmitteckert, J. Kroha 
and P. W\"{o}lfle {\em Physica C}{\bf 235--240}, 2287 (1994).
\bibitem{nrg.ref} K. G. Wilson, {\em Rev. Mod. Phys. }{\bf 47}, 773
(1975); H. B. Krishnamurthy, J. W. Wilkins \&
K. G. Wilson, {\em Phys. Rev. }{\bf B21}, 1044 (1980).
\bibitem{costi.94c} T. A. Costi, A. C. Hewson and V. Zlati\'{c} {\em J.
Phys.: Cond. Matt.} {\bf 6}, 2251 (1994); T. A. Costi, A. C. Hewson, 
{\em J. Phys.: Cond. Matt.} {\bf 5}, 361 (1993); T. A. Costi, A. C. Hewson, 
{\em Phil. Mag.} B{\bf 65}, 1165 (1992). 
\bibitem{anderson.67} P. W. Anderson, {\em Phys. Rev. Lett}
{\bf 18}, 1049 (1967).
\bibitem{schotte.69} K. D. Schotte and U. Schotte, {\em Phys. Rev.}
{\bf 185}, 509 (1969).
\bibitem{read.85} N. Read, {\em J. Phys. C}{\bf 18}, 2651 (1985).
\bibitem{mh.88} B. Menge and E. M\"{u}ller--Hartmann, Z. Phys. {\bf
B73}, 225 (1988).
\bibitem{gruneberg.91} J. Gruneberg and H. Keiter,
{\em Physica B}{\bf 171}, 39 (1991).
\bibitem{anders.vertex} F. Anders and N. Grewe, Europhys. Lett. {\bf
26}, 551 (1994); F. Anders, J. Phys. Cond. Mat. {\bf 7}, 2801 (1995).
\bibitem{kroha.tmatrix} J. Kroha, P. J. Hirschfeld, K. A. Muttalib 
and P. W\"{o}lfle, {\em Solid State Commun.}{\bf 83}, 1003 (1992).
\bibitem{kroha.privatecommun.95} J. Kroha, in preparation (1995).
\end{references}
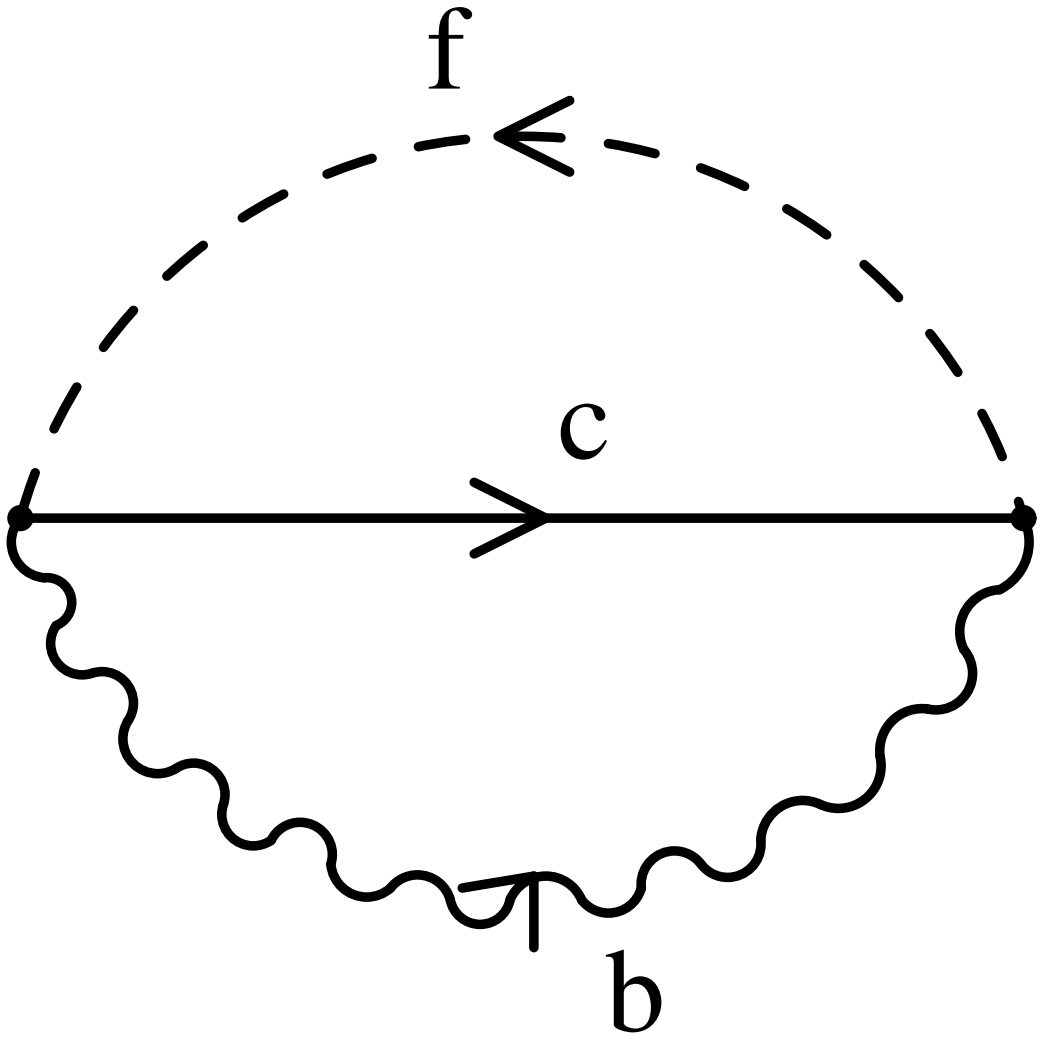
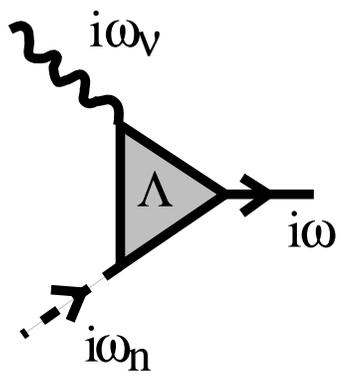
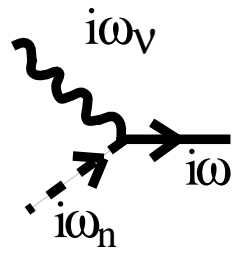
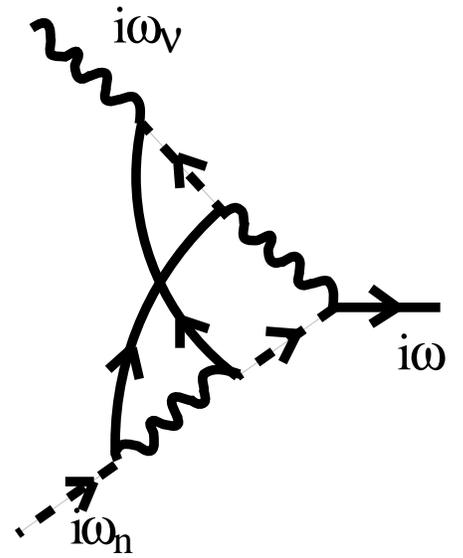
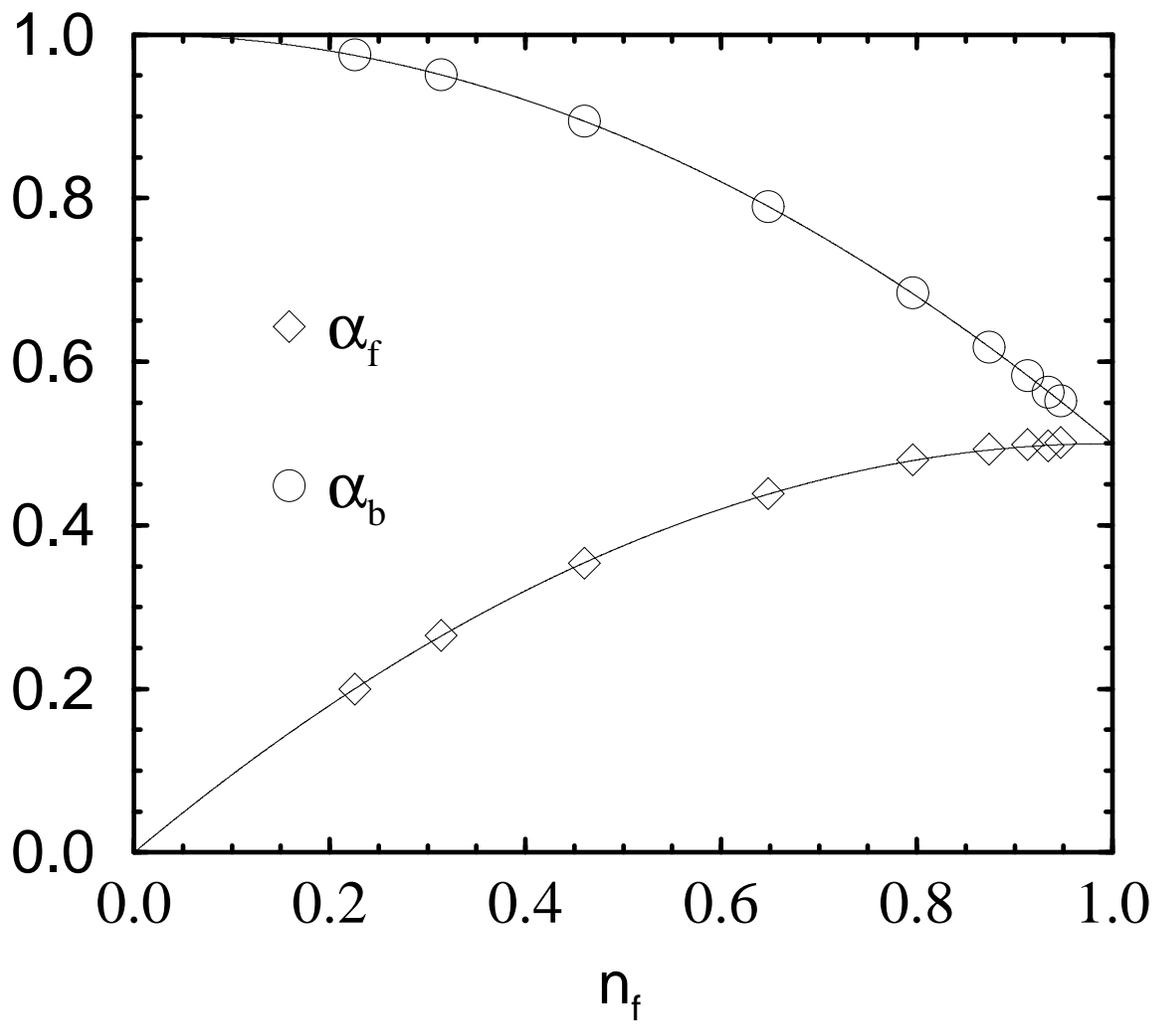
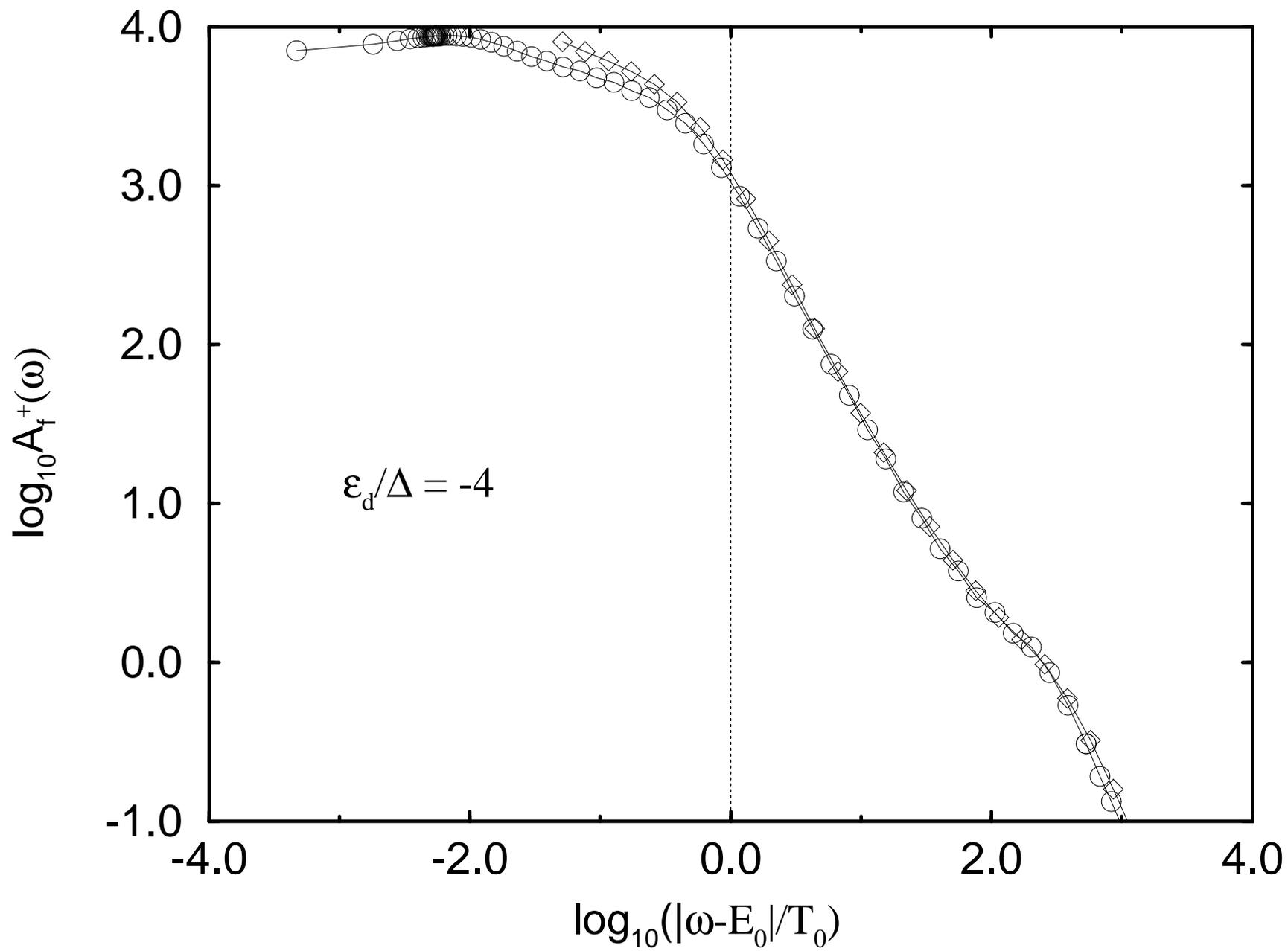
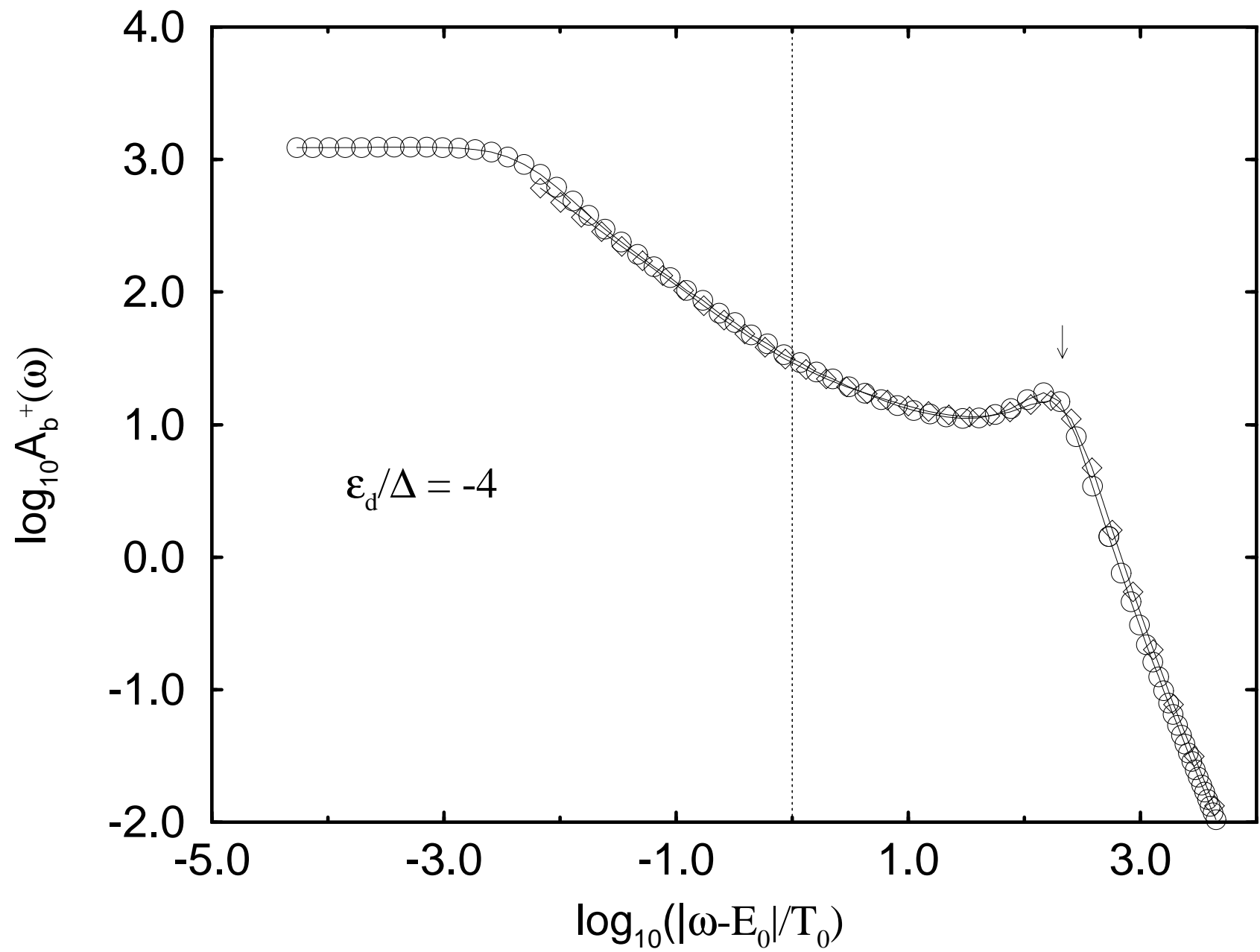
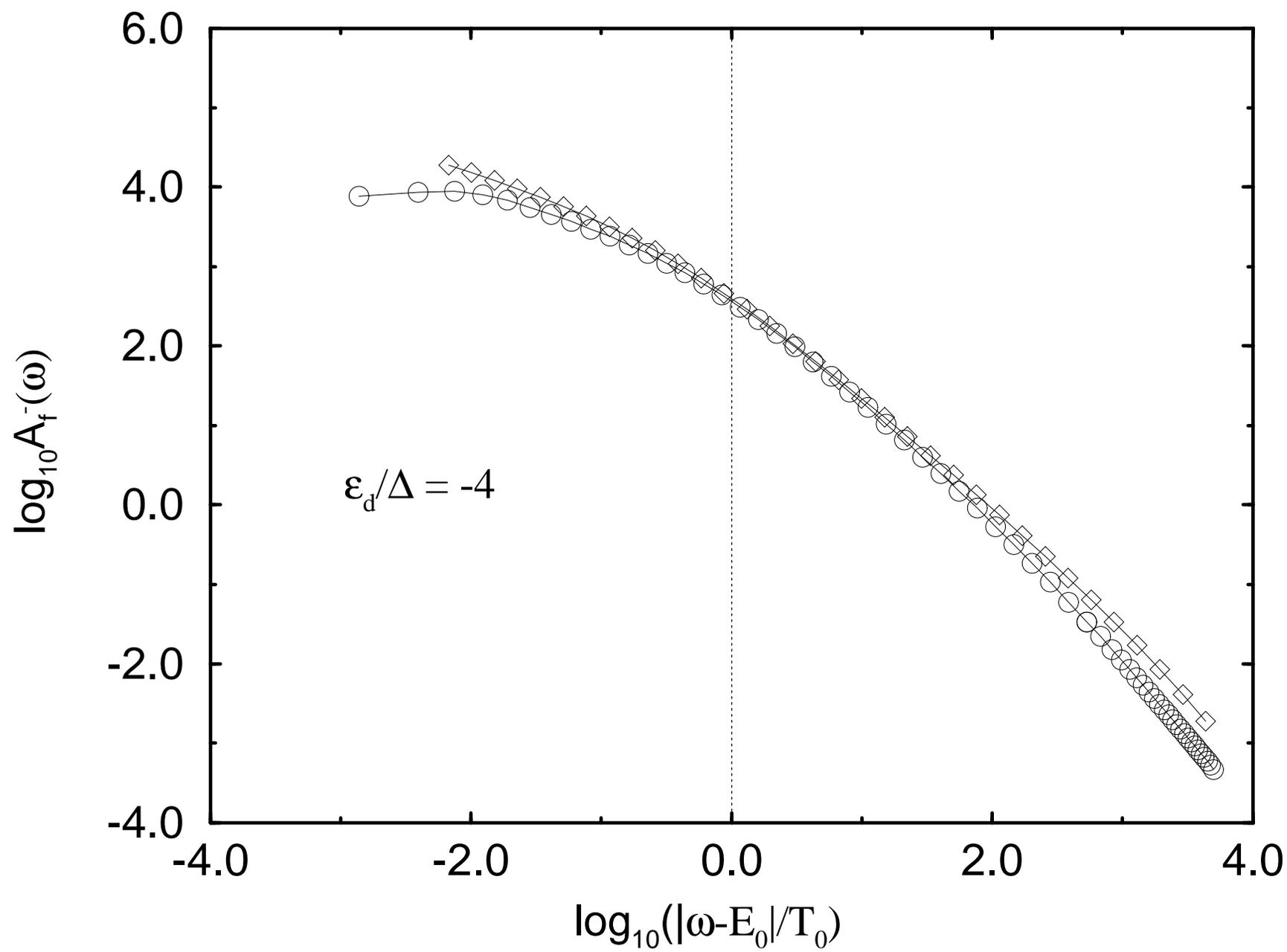
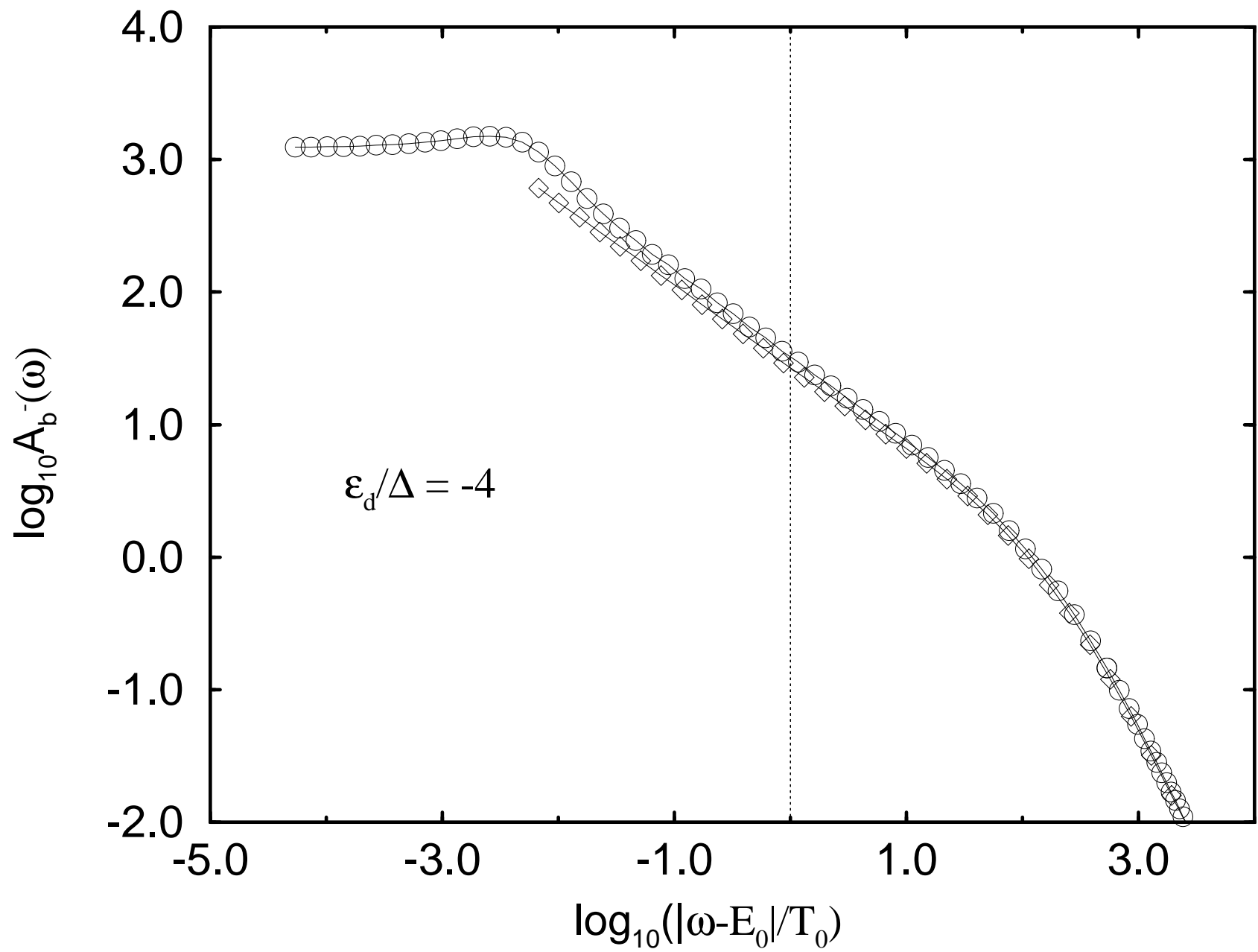
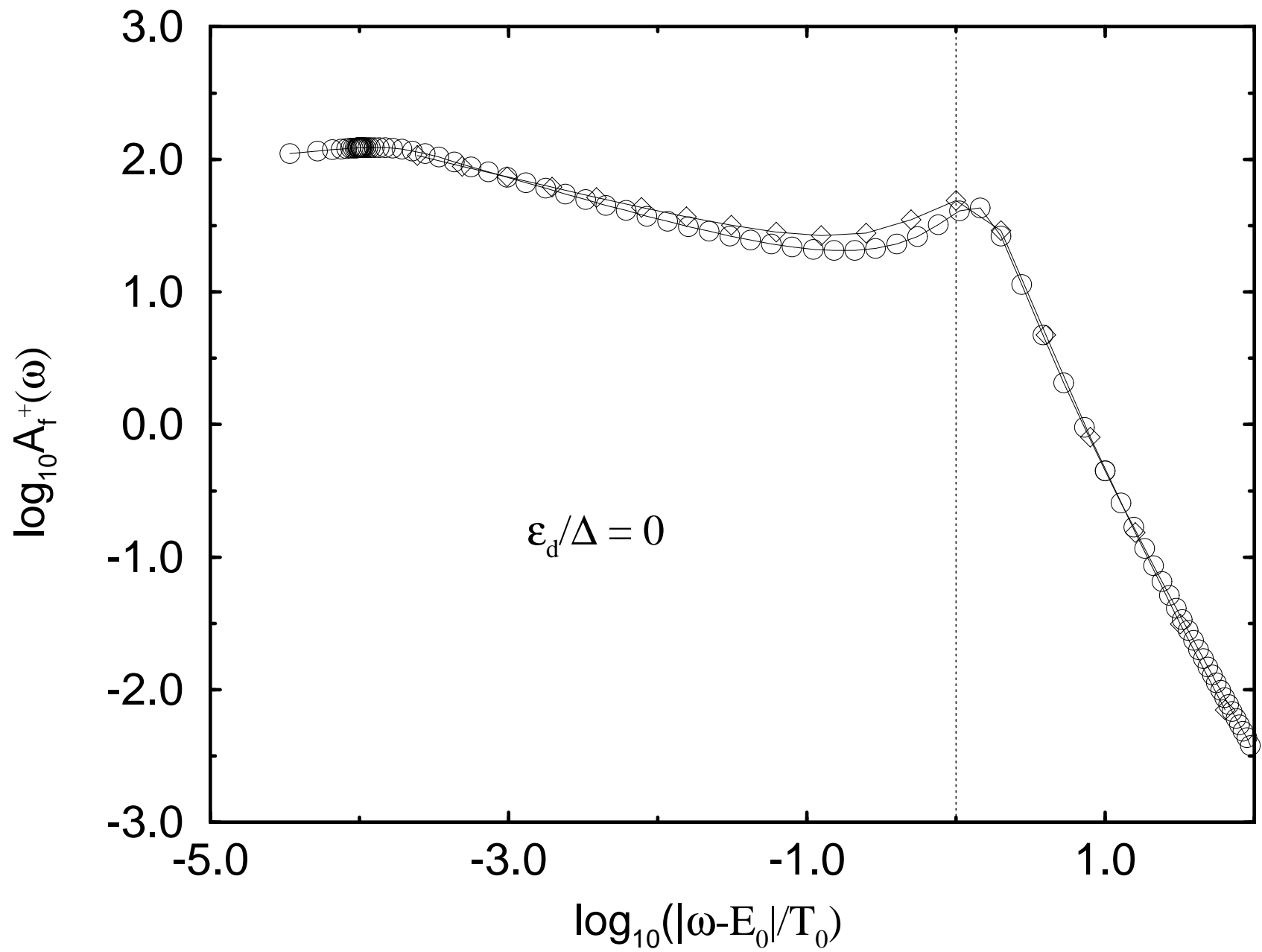
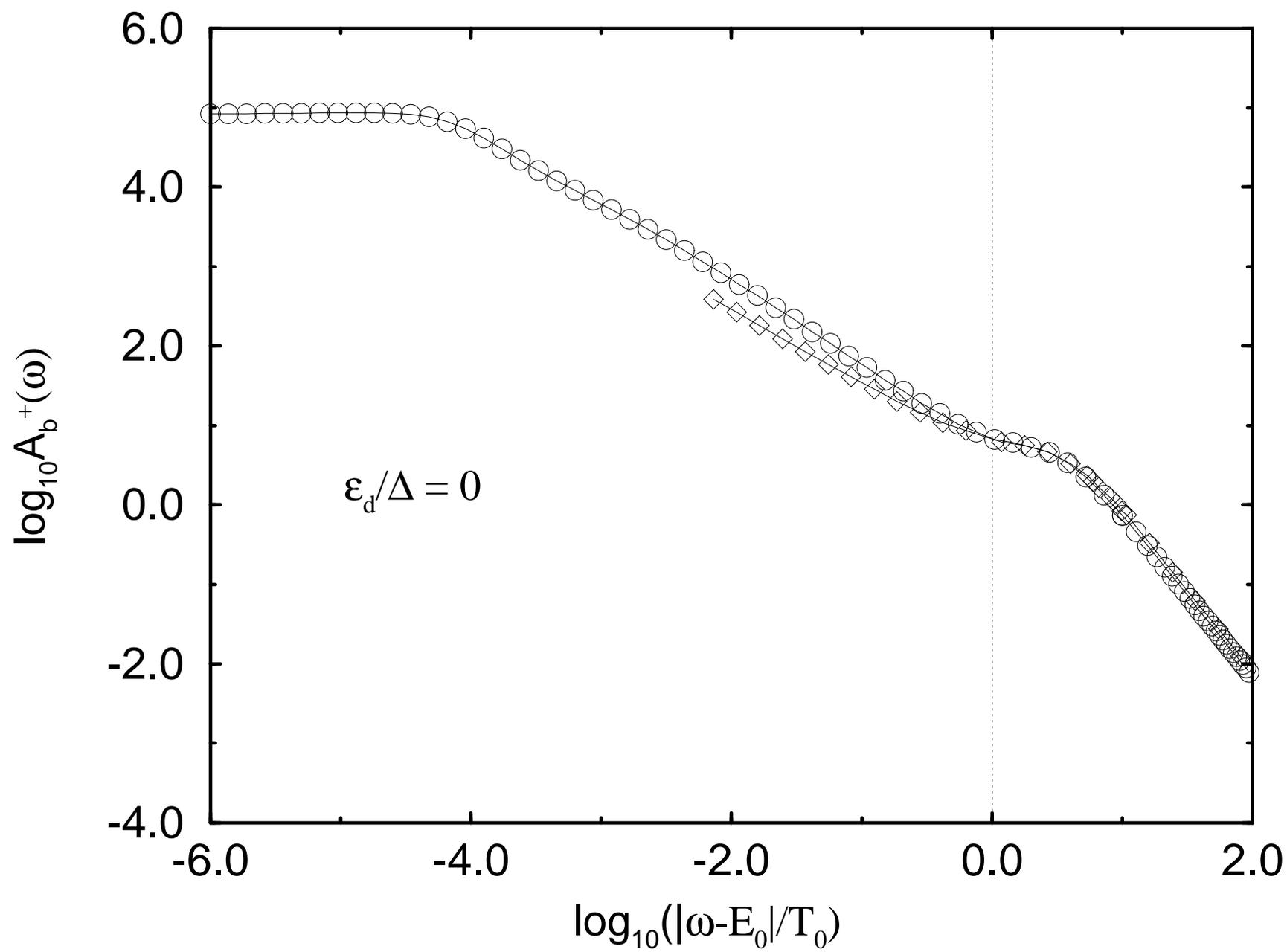
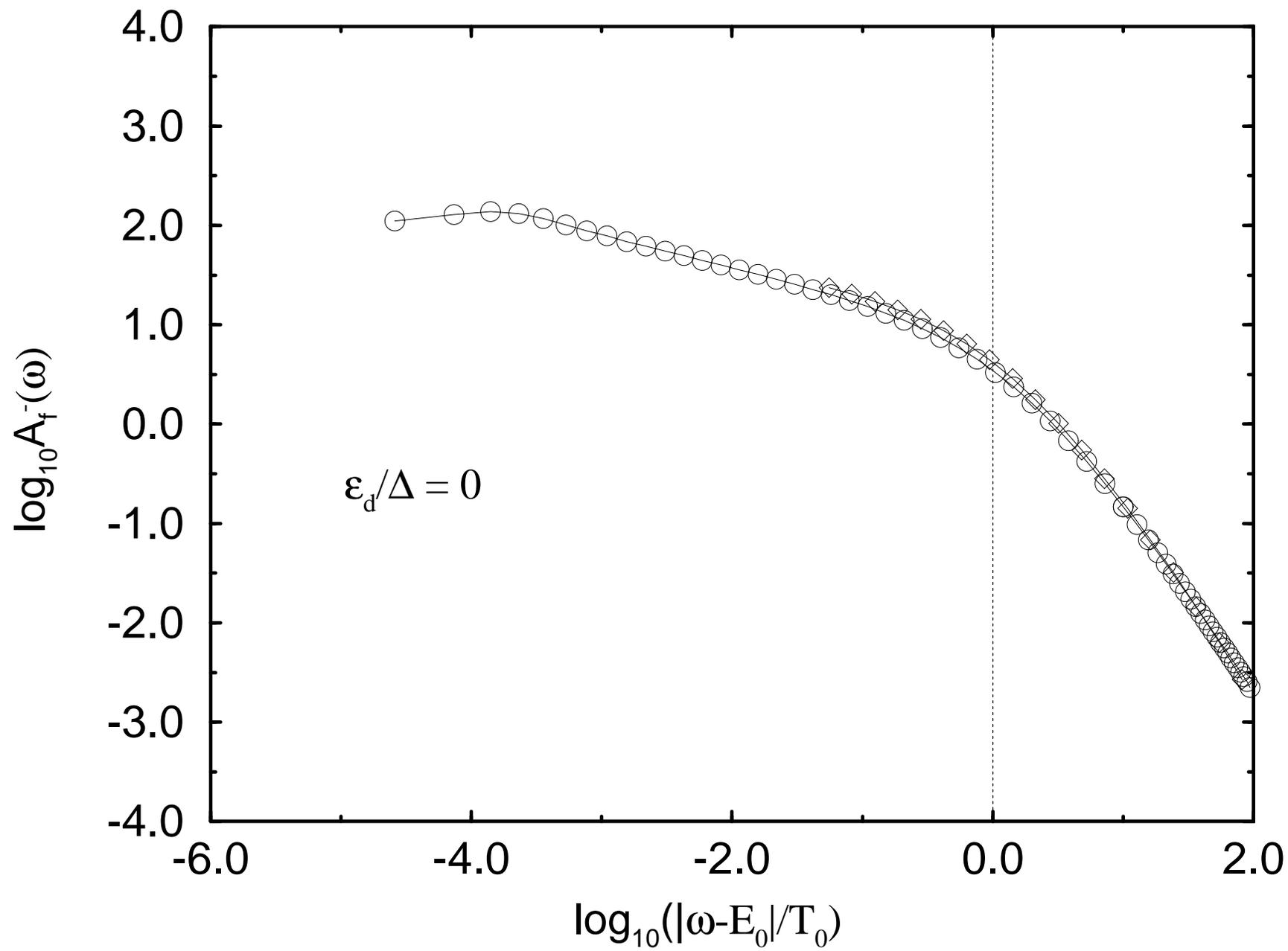
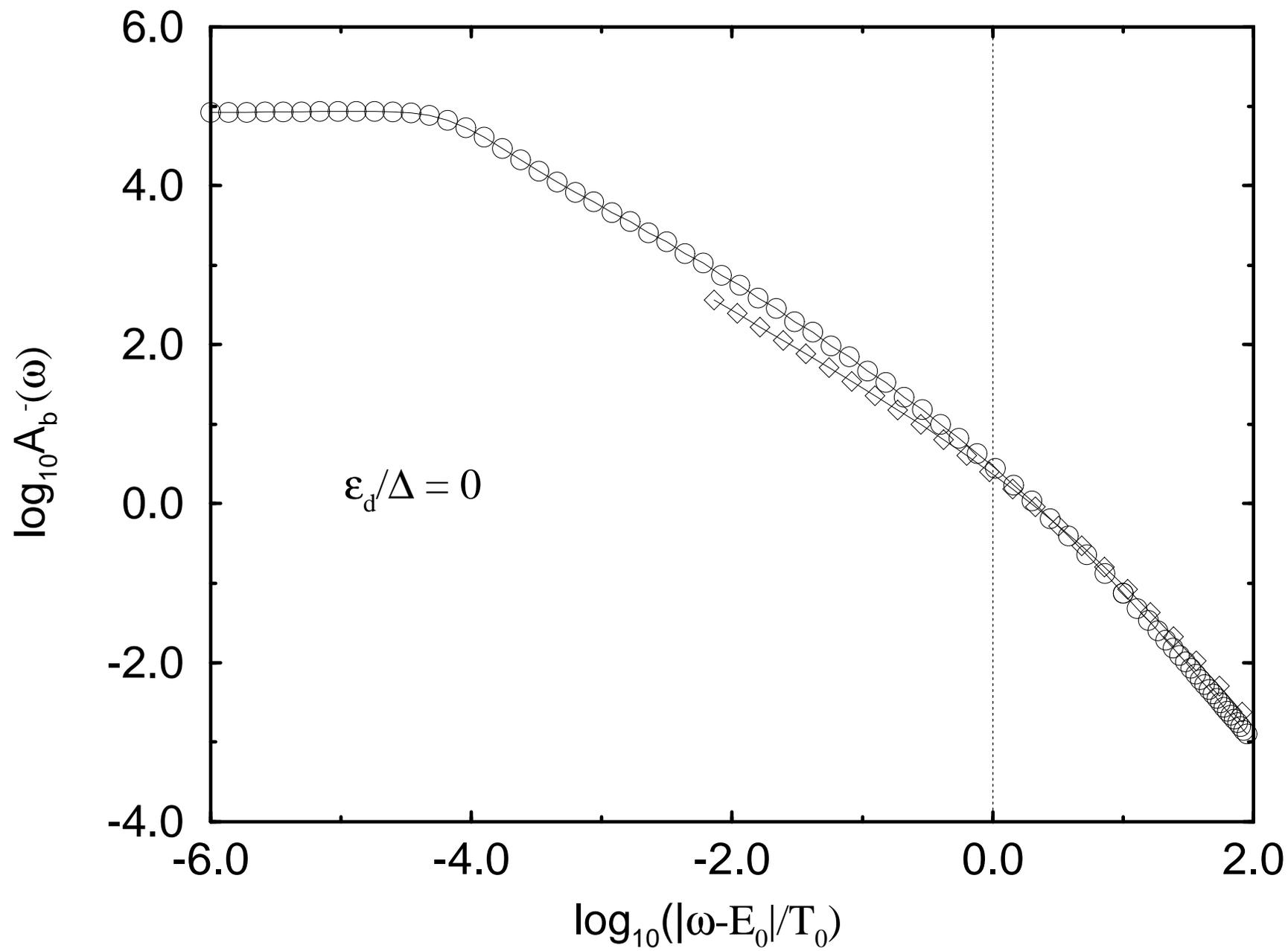
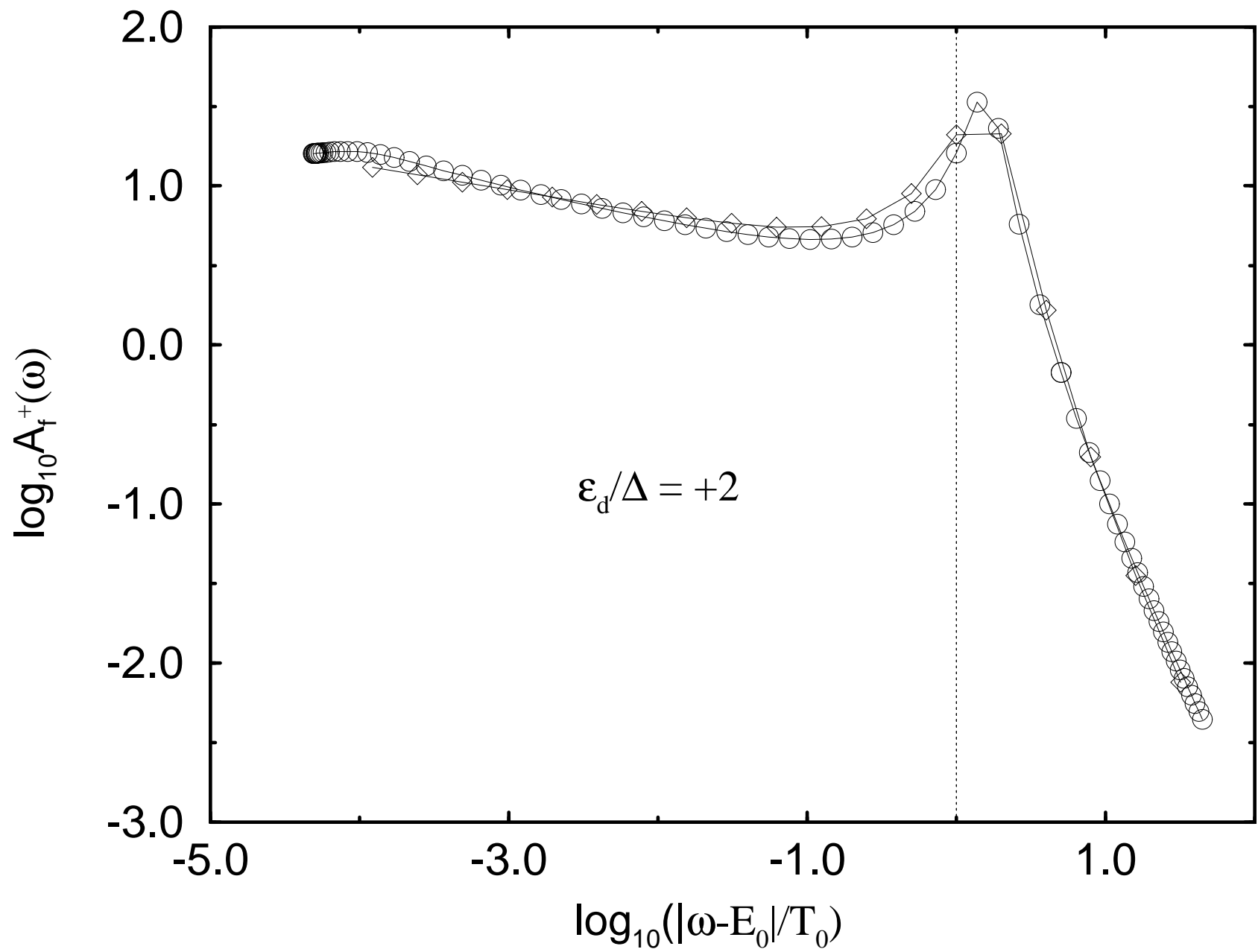
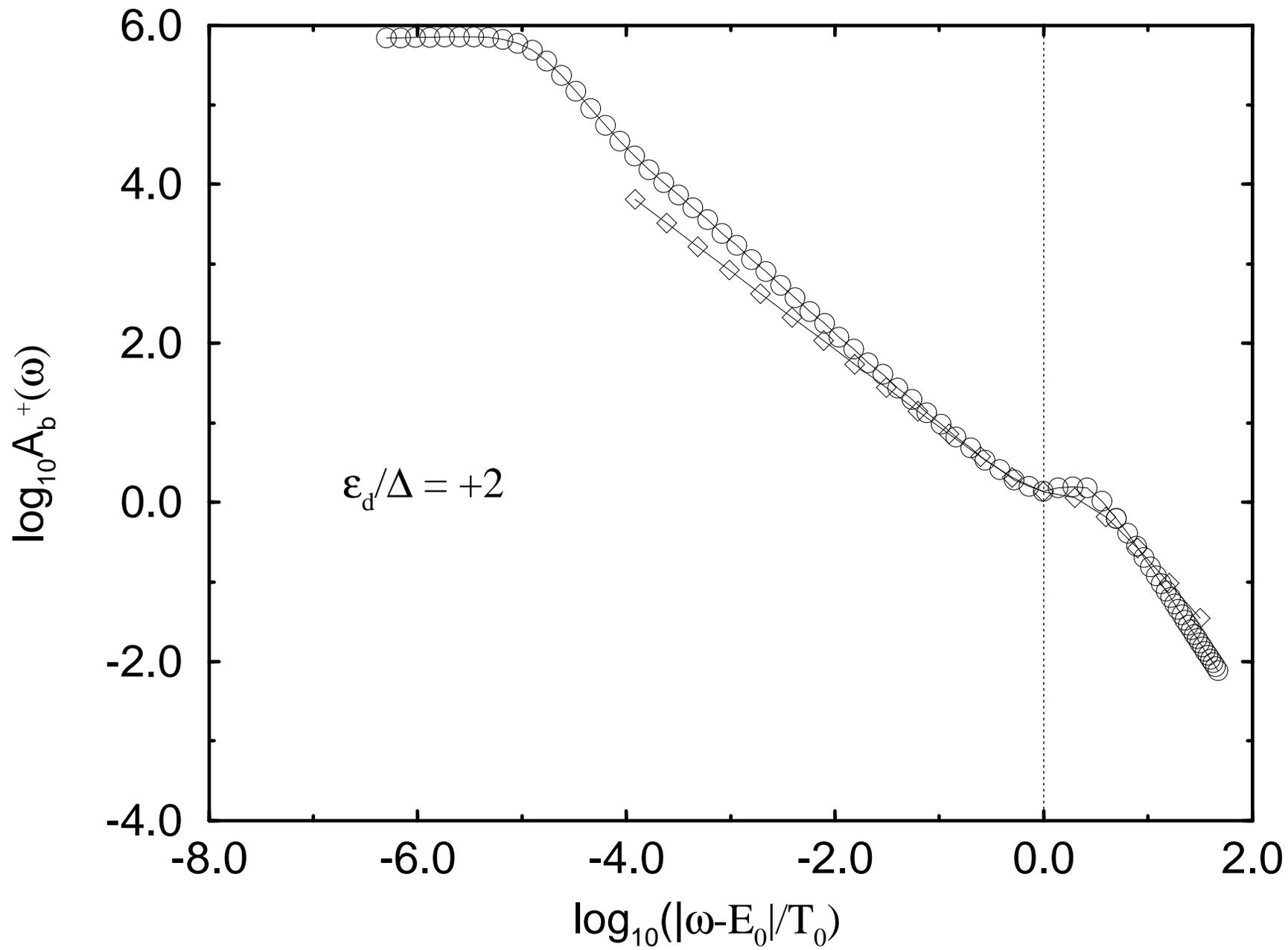
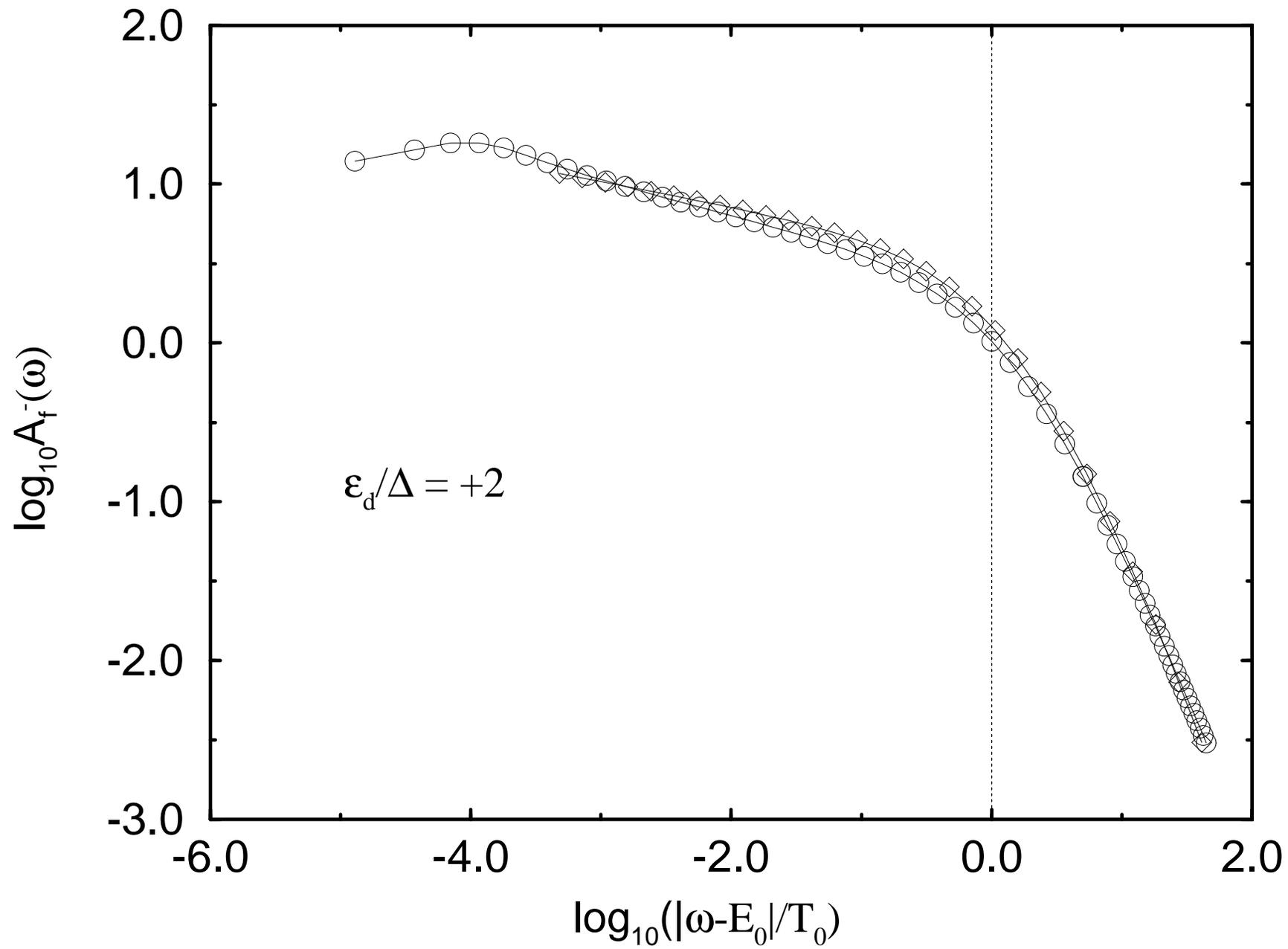
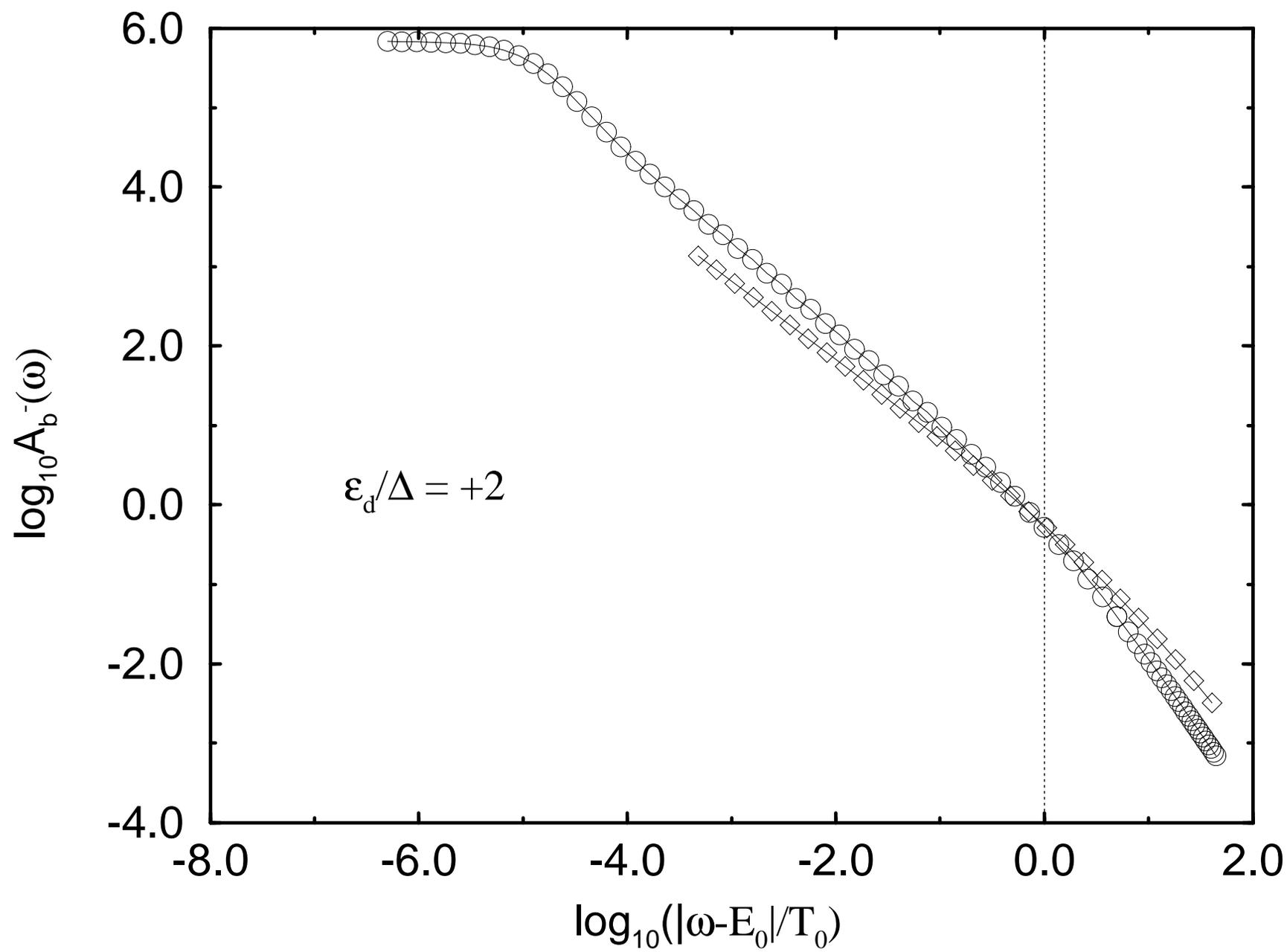
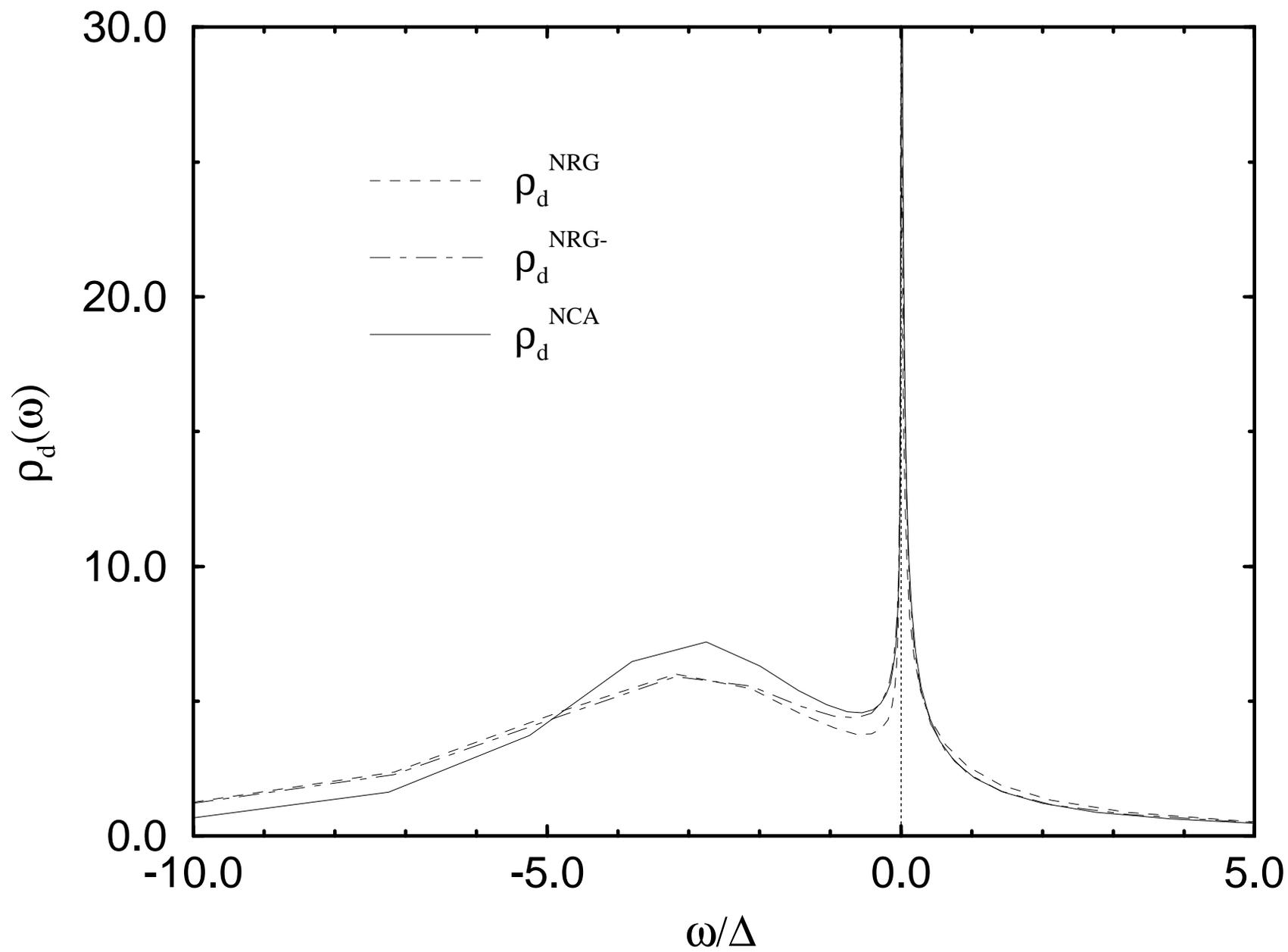
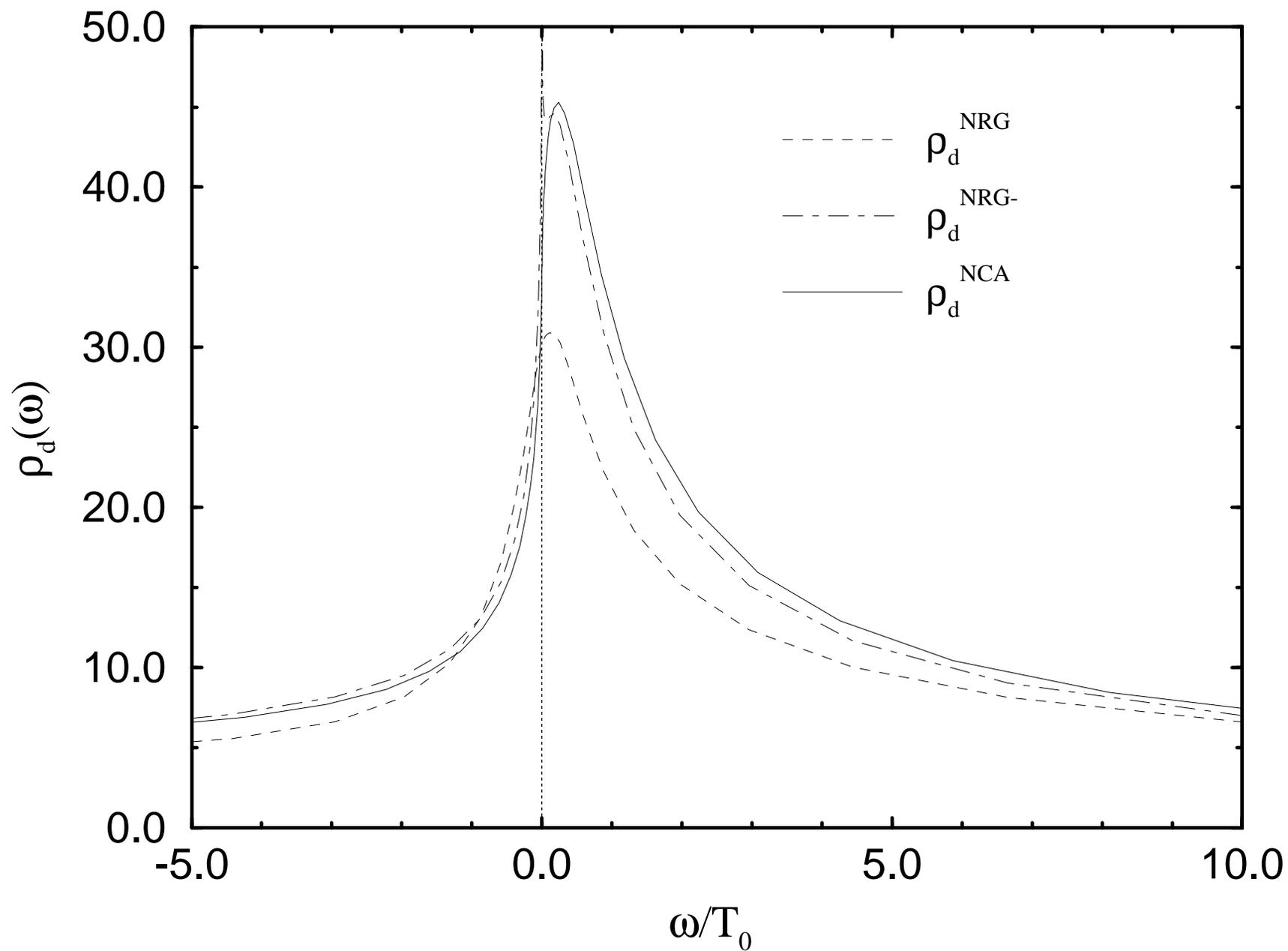
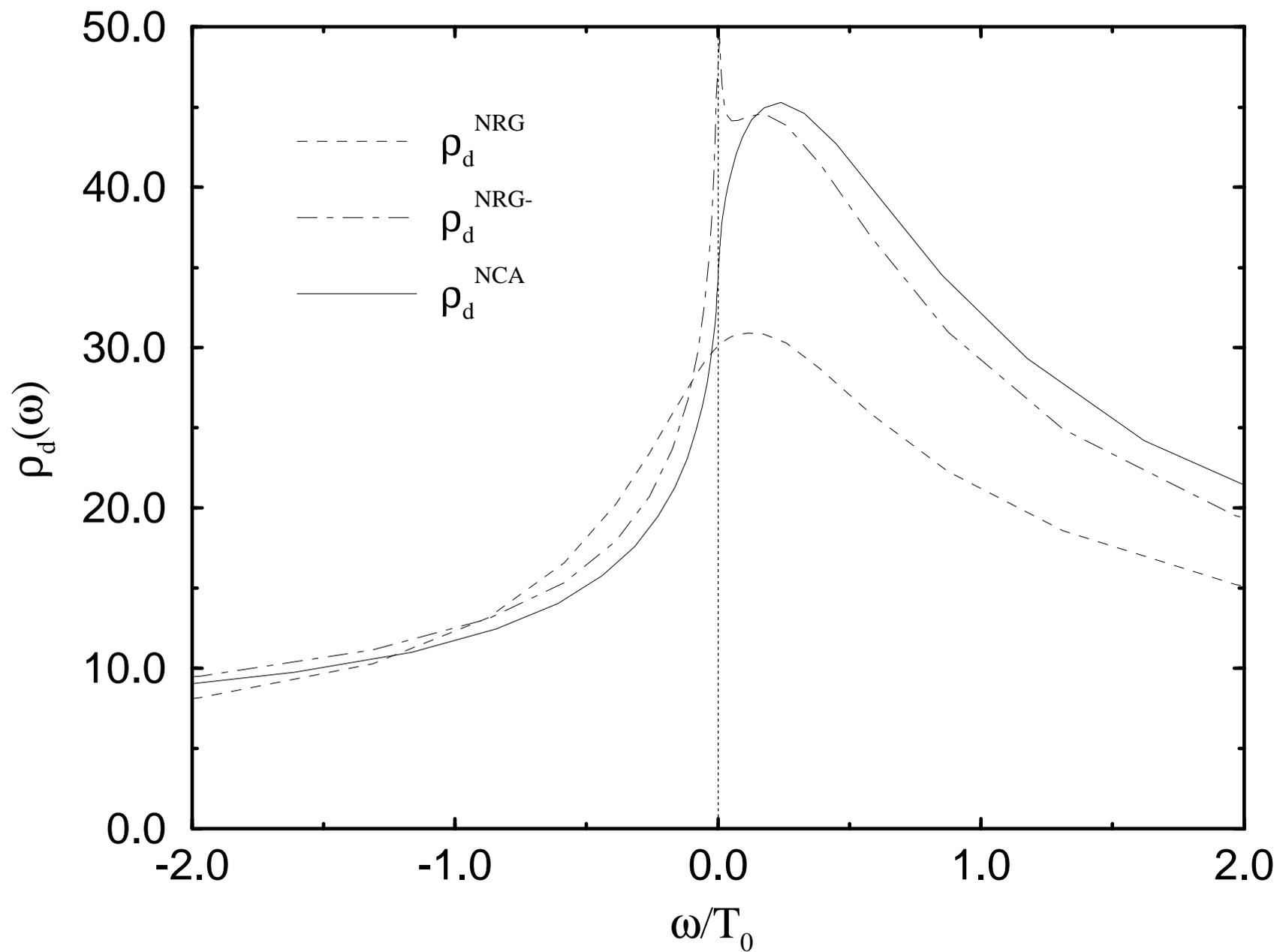
\end{document}